\documentclass[11pt]{article}

\usepackage{geometry}
\usepackage{amsfonts,amssymb,amsmath,amsthm}
\usepackage{graphicx}
\usepackage{color}

\newcommand{\ncd}{\newcommand}
\ncd{\QCcns}{$QC_{\cal{C}}$}
\ncd{\QCc}{$QC_{\cal{C}}\;$}

\begin{document}

\title{Topological fault-tolerance in cluster state quantum computation}

\author{R. $\mbox{Raussendorf}^1$, J. $\mbox{Harrington}^2$ and
  K. $\mbox{Goyal}^3$
\vspace{1.5mm}\\
$\mbox{}^1${\em{\small{Perimeter Institute for Theoretical Physics,
  Waterloo, ON, M6P 1N8, Canada}}}\\
$\mbox{}^2${\em{\small{Applied Modern Physics, MS D454, Los Alamos National Laboratory,
Los Alamos, NM 87545, USA}}}\\
$\mbox{}^3${\em{\small{Institute for Quantum Information, California
      Institute of Technology,
Pasadena, CA 91125, USA}}}
}

\maketitle

\begin{abstract}
   We describe a fault-tolerant version of the one-way quantum
   computer using a cluster state in three spatial
   dimensions. Topologically protected quantum gates are realized
   by choosing appropriate boundary conditions on the cluster. We
   provide equivalence transformations for these boundary conditions
   that can be used to simplify fault-tolerant circuits and to derive
   circuit identities in a topological manner.
   The spatial dimensionality of the scheme can be reduced to two by
   converting one spatial axis of the cluster into time. The error threshold
   is 0.75\% for each source in an error model with preparation,
   gate, storage and measurement errors. The operational overhead is
   poly-logarithmic in the circuit size.
\end{abstract}


\section{Introduction}
\label{intro}

The threshold theorem for fault-tolerant quantum computation
\cite{TT1,TT2,TT3,TT4} has established the fact that
large quantum computations can be performed with arbitrary accuracy,
provided that the error level of the elementary components of the quantum
computer is below a certain threshold. It now becomes important to
devise methods for error correction which yield a
high threshold,  are robust against variations of the error model,
and can be implemented with small operational overhead.
An additional desideratum is a simple architecture for the quantum
computer, such as requiring no long-range interaction.

The one-way quantum computer provides a method to do this
\cite{RHG, RH}, which we describe in detail below. We obtain an error threshold estimate
of 0.75\% for each source in an error model with preparation, gate,
storage and measurement errors, with a poly-logarithmic multiplicative
overhead in the
circuit size $\Omega$ ($\sim \ln^3 \Omega$). It shall be noted that we
achieve this threshold in a 2-dimensional geometry, only requiring
nearest-neighbor translation-invariant Ising interaction. This is
relevant for experimental realizations based on matter qubits such as
cold atoms in optical lattices \cite{OL1,OL2} and two-dimensional ion
traps \cite{ITr}, or stationary qubits in quantum dot systems \cite{Qdot}
and arrays of superconducting qubits \cite{SCQ}. Geometric constraints
are no major concern for fault-tolerant quantum computation with photonic qubits \cite{Browne, Nielsen}.

The highest known threshold estimate, for a setting without geometric
constrains, is $3\times 10^{-2}$ \cite{Kn2}.
Fault-tolerance is more difficult to achieve in architectures where
each qubit can only interact with other qubits in its immediate
neighborhood. A recent fault-tolerance threshold for
a two-dimensional lattice of qubits with only local and
nearest-neighbor gates is $1.9 \times 10^{-5}$ \cite{Svo}. We note
that since the initial work of \cite{Kit1} a number of distinct approaches
to topological fault-tolerance emerging in lattice systems are being
pursued; See \cite{Bombin1, Bombin2, Pach1, Pach2}.

\begin{figure}
  \begin{center}
    \includegraphics[width=14cm]{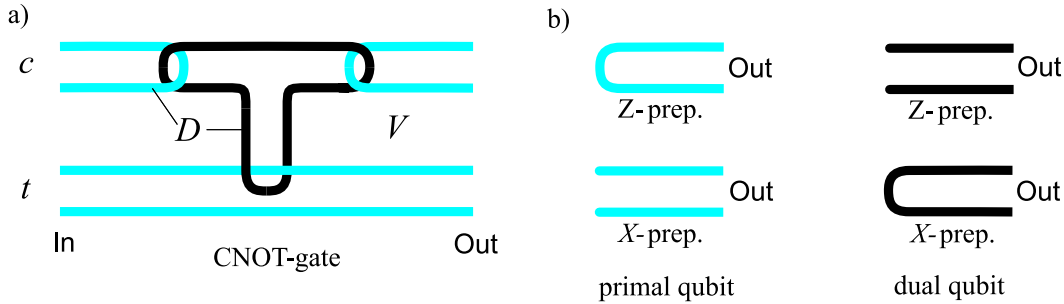}
    \caption{\label{tg} Topologically protected gates.}
  \end{center}
\end{figure}

The key element of our method is based on topological tools that become
available when  the dimensionality of the cluster is increased from two
to three. In 3D, we combine the universality
already found in 2D cluster states \cite{QCc} with the topological
error-correcting capability of Kitaev's toric code \cite{Kit1}. Then,
a one-dimensional sub-structure of the cluster is ``carved out'' by
performing local $Z$-measurements. This leaves us with a non-trivial
cluster topology in which a fault-tolerant quantum circuit is embedded.
Fig.~\ref{tg} displays topologically protected gates that can be
constructed in this manner.

This paper is organized as follows. In the remainder of this section
we introduce the necessary terminology for discussion of the
fault-tolerant \QCcns. In Section~\ref{TQ} we answer the question of
``Why can we
perform non-abelian gates with surface codes?'' and describe
topological transformation rules on the cluster. We subsequently use
these rules to
simplify topological circuits. In Section~\ref{USG} we complete the
universal set of gates. In Section~\ref{2D} we describe the mapping
from a 3D cluster state to a two-dimensional physical system plus
time. Sections~\ref{FTT} and \ref{OH} address
the fault-tolerance threshold and overhead, respectively. We
conclude with a summary and outlook in Section~\ref{OL}.\medskip

\begin{figure}
  \begin{center}
    \includegraphics[width=8.6cm]{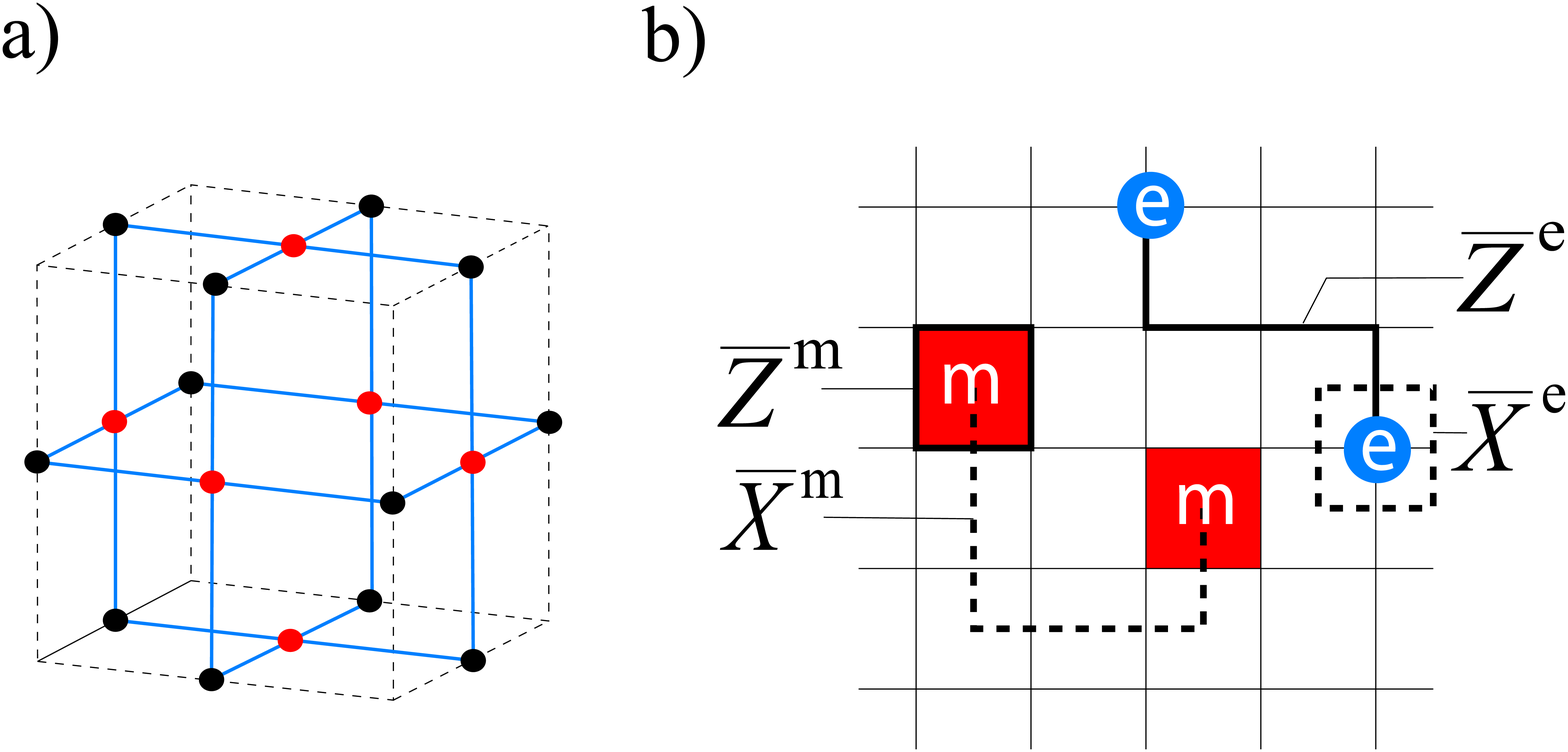}
    \caption{\label{LattDef}Lattice definitions. a)
      Elementary cell of
    the cluster lattice ${\cal{L}}$. 1-chains of
    ${\cal{L}}$ (dashed lines), and graph edges (solid lines). b) A
    pair of electric (``e'') or magnetic
    (``m'') holes in the code
    plane each support an encoded qubit. $\overline{Z}^{e/m}$
    and $\overline{X}^{e/m}$ denote the encoded Pauli operators $Z$ and
    $X$, respectively.}
  \end{center}
\end{figure}

Before we can start our discussion of the fault-tolerance
properties of the \QCcns, we need to introduce some necessary
notation. This has been done before in \cite{RHG, RH}. We include a
short introduction here to make our presentation
self-contained. Consider a cluster state
$|\phi\rangle_{\cal{L}}$ on a lattice ${\cal{L}}$ with elementary cell
as displayed in Fig.~\ref{LattDef}a. Qubits are located at the center of faces
and edges of ${\cal{L}}$. The lattice ${\cal{L}}$ is
subdivided into three regions $V$, $D$ and $S$.
Each region has its purpose, shape and specific measurement basis
for its qubits. The qubits in $V$ are measured in the $X$-basis, the
qubits in $D$ in the $Z$-basis, and the qubits in $S$ in either of
the eigenbases $(X \pm Y)/\sqrt{2}$. $V$ fills up most of the cluster.
$D$ is composed of thick line-like structures, named {\it
  defects}. $S$ is composed
of well-separated qubit locations interspersed among the defects.
As described in greater detail below, the cluster region $V$ provides
topological
error correction, while regions $D$ and $S$ specify the Clifford and
non-Clifford parts of a quantum algorithm, respectively.

We can break up this measurement pattern into gate
simulations by establishing the following correspondence:
$\mbox{\em{quantum gates}} \leftrightarrow
\mbox{\em{quantum correlations}} \leftrightarrow
\mbox{\em{surfaces}}$.
The first part of this correspondence has been established in
\cite{RBB03}. For the second part homology comes into play.
The correlations of $|\phi\rangle_{\cal{L}}$, i.e., the stabilizers,
can be identified with 2-chains (surfaces) in ${\cal{L}}$, while errors
map to 1-chains (lines). Homological equivalence of the chains
implies physical equivalence of the corresponding operators \cite{RHG}. This
correspondence is key to the presented scheme. Gates are specified by
a set of surfaces with input and  output boundaries, and syndrome
measurements correspond to closed surfaces (having no boundary).

${\cal{L}}$ is regarded as a chain complex,
${\cal{L}}=\{C_3,C_2,C_1,C_0\}$. It has a dual
${\overline{\cal{L}}}=\{\overline{C}_3,\overline{C}_2,\overline{C}_1,
\overline{C}_0\}$ whose cubes $\overline{c}_3 \in \overline{C}_3$
map to sites $c_0 \in C_0$ of ${\cal{L}}$, whose faces
$\overline{c}_2 \in \overline{C}_2$ map to edges $c_1 \in C_1$ of
${\cal{L}}$, etc. The chains have coefficients in $\mathbb{Z}_2$. One
may switch back and forth between
${\cal{L}}$ and $\overline{\cal{L}}$ by a duality transformation
$\mbox{}^*(\;)$.  ${\cal{L}}$ and $\overline{\cal{L}}$ are each equipped with
a boundary map $\partial$, where $\partial \circ \partial = 0$.

Operators may be associated with chains as follows. Suppose that for each
qubit location $a$ in a chain $c$, $a \in \{c\}$, there exists an operator
$\Sigma_a$, and $[\Sigma_a,\Sigma_b]=0$ for all $a,b \in
\{c\}$. Then, we define $\Sigma(c) := \prod_{a \in \{c\}} \Sigma_a$.
Cluster state correlations (i.e. stabilizers) are associated with
2-chains. For the considered lattice, all
elements in the cluster state stabilizer take the form
$K(c_2)K(\overline{c}_2)$ with $c_2 \in C_2$, $\overline{c}_2 \in
\overline{C}_2$, and
\begin{equation}
  K(c_2)= X(c_2)Z(\partial c_2),\;\; K(\overline{c}_2)=
  X(\overline{c}_2)Z(\partial \overline{c}_2).
\end{equation}
Only those stabilizer elements compatible with the local measurement
scheme are useful for information processing. In particular, they need to
commute with the measurements in $V$ and $D$,
\begin{equation}
  \label{CompatCond}
  \begin{array}{rclcr}
    {[K(c_2)K(\overline{c}_2), X_a]}  &=& 0, && a \in V, \\
    {[K(c_2)K(\overline{c}_2), Z_b]}  &=& 0, && b \in D.
  \end{array}
\end{equation}
Due to the presence of a primal lattice ${\cal{L}}$ and a dual lattice
$\overline{\cal{L}}$, it is convenient to subdivide the sets $V$ and $D$
into primal and dual subsets. Specifically,
$V = V_p \cup V_d$, with $V_p \subset \{C_2\}$, $V_d \subset
\{\overline{C}_2\}$, and $D = D_p \cup D_d$, with $D_p \subset
\{C_1\}$, $D_d \subset \{\overline{C}_1\}$. With these notions
introduced the compatibility condition (\ref{CompatCond}) may be
expressed directly in terms of the chains $c_2$ and
$\overline{c}_2$. If $K(c_2)$ and
$K(\overline{c}_2)$ have support in $V \cup D$ only then
Eq. (\ref{CompatCond}) is equivalent to
\begin{equation}
  \{\partial c_2\} \subset D_p, \;\; \{\partial \overline{c}_2\} \subset D_d.
\end{equation}

\paragraph{The \QCc and surface codes.} We need to specify the
encoding of logical qubits before explaining the encoded gates.
For this purpose, let us single out one spatial direction on the cluster
as `simulating time.' The perpendicular 2D slices provide space for a
quantum code. The code which fills this plane after the mapping of the
three dimensional lattice ${\cal{L}}$ onto a 2+1 dimensional one is
the {\em{surface code}} \cite{Kit2}.

The number of qubits which can be encoded in such a code
depends solely on the surface topology. Here we consider
a plane with pairs of either electric or magnetic holes.
See Fig.~\ref{LattDef}b. A magnetic hole is a plaquette $f$ where
the associated stabilizer generator $S_\Box(f)=Z(\partial{f})$ is
{\em{not enforced}} on the code space, and an electric hole is a site
$s$ where the associated stabilizer $S_+(s)=X(\partial \mbox{ }^\# s)$
is {\em{not enforced}} on the code space (``$\#$'' denotes the duality
transformation in 2D). Each hole is the intersection of a defect
strand with a constant-time slice.

A pair of holes supports a qubit. For a pair of magnetic holes $f,
f'$, the encoded spin flip operator is $\overline{X}^m =
X(\overline{c}_1)$, with $\{\partial \overline{c}_1\} =\{\mbox{}^\#f,
\mbox{}^\#f^\prime \}$, and the
encoded phase flip operator is $\overline{Z}^m= Z(c_1)$, with $c_1
\cong \partial f$ or $c_1 \cong \partial f^\prime$. The operator
$Z(\partial f + \partial f^\prime)$ is in the code stabilizer
${\cal{S}}$,
\begin{equation}
  \label{SGm}
  Z(\partial f + \partial f^\prime) \in {\cal{S}}.
\end{equation}
For a pair of electric holes $s, s^\prime$ we have $\overline{X}^e =
X(\overline{c}_1^\prime)$, with $\overline{c}_1^\prime \cong \partial
\mbox{}^\#s$, $\overline{Z}^e = Z(c_1)$, with $\{\partial c_1\} = \{s,
s^\prime \}$, and
\begin{equation}
\label{SGe}
X(\partial \mbox{}^\#s+ \partial
\mbox{}^\#s^\prime) \in {\cal{S}}.
\end{equation}

\paragraph{The simplest gate.} Here we illustrate the
relation between quantum gates, quantum correlations and correlation
surfaces (2-chains). We choose the simplest possible example: the identity
gate.

The identity operation is realized by two parallel strands of defect
of the same type. We consider a block shaped cluster ${\cal{C}}
\subset {\cal{L}}$ for
the support of the identity gate. One of the spatial
directions on the cluster is singled out as `simulated time.' The two
perpendicular
slices of the cluster at the `earliest' and `latest' times represent
the code surfaces $I$ and $O$ for the encoded qubit, with $I,O \subset
\{C_1\}$ being an integer number of elementary cells apart. As before,
we ask which
cluster state correlations $K(c_2)$, $K(\overline{c}_2)$ are
compatible with the local measurements in ${\cal{C}}\backslash (I \cup
O)$. With the additional regions $I$ and $O$ present, the condition
(\ref{CompatCond}) turns into
\begin{equation}
  \label{CCC}
  \begin{array}{rclcrcl}
    \{c_2\} &\subset& V_p, && \{\partial c_2\} &\subset& D_p \cup I \cup
  O,\\
  \{\overline{c}_2\} &\subset& V_d \cup I \cup O, &&  \{\partial
  \overline{c}_2\} &\subset& D_d.
  \end{array}
\end{equation}
Surfaces of primal correlations compatible with the local measurements in
${\cal{C}}\backslash (I \cup O)$ can stretch through the cluster
region $V$ and end in the primal defects and the input- and output
regions. They cannot end in a dual defect. Surfaces of dual
correlations can stretch through $V$, $I$
and $O$, and end in dual defects. They cannot end in primal
defects\footnote{The asymmetry between primal
  and dual 2-chains in Eq. (\ref{CCC}) arises because $I$ and
$O$ are chosen subsets of $C_1$. Physically speaking, we choose the
sub-cluster ${\cal{C}}$ such that it consists of intact
cells of the primal lattice ${\cal{L}}$ at the front and back. The
cells of the dual lattice $\overline{\cal{L}}$ are
then cut in half at the front and back of ${\cal{C}}$.}.

\begin{figure}
  \begin{center}
    \includegraphics[width=7cm]{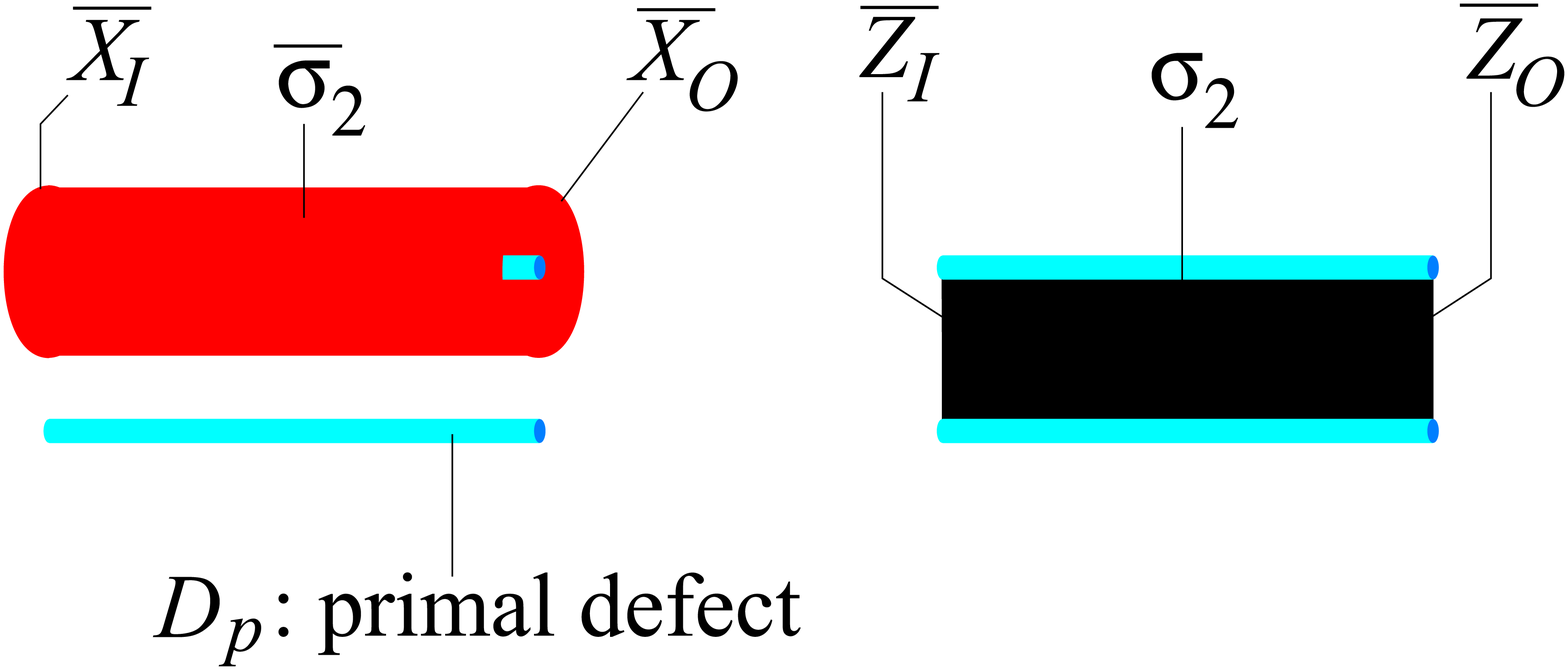}
    \caption{\label{IdGate}Correlation surfaces $\sigma_2$,
      $\overline{\sigma}_2$ for the
      identity gate on a primal qubit.}
  \end{center}
\end{figure}

We now consider the identity gate on the primal qubit, mediated by a
pair of primal defects. The relevant primal and dual correlation
surfaces are displayed in Fig.~\ref{IdGate}, and we
denote these special surfaces by $\sigma_2$ and $\overline{\sigma}_2$.
Before the local measurement of the qubits in ${\cal{C}}\backslash (I \cup
O)$ the cluster state $|\phi\rangle_{\cal{C}}$ obeys
$K(\sigma_2)|\phi\rangle_{\cal{C}} = K(\overline{\sigma}_2)
|\phi\rangle_{\cal{C}} = |\phi\rangle_{\cal{C}}$. Note that
$\left. K(\sigma_2)\right|_{I\cup O}=\overline{Z}_I\otimes
\overline{Z}_O$ and $\left. K(\overline{\sigma}_2)\right|_{I\cup
  O}=\overline{X}_I\otimes \overline{X}_O$. The ``$\overline{\cdot}$''
refers to encoding with the surface code displayed in Fig.~\ref{LattDef}.
Thus, for the state
$|\psi\rangle_{I \cup O}$ after the measurements in
${\cal{C}}\backslash (I \cup O)$, $\overline{Z}_I\otimes
\overline{Z}_O |\psi\rangle_{I \cup O} = \pm
|\psi\rangle_{I \cup O}$ and $\overline{X}_I\otimes
\overline{X}_O|\psi\rangle_{I \cup O} =\pm|\psi\rangle_{I \cup O}$. This
is the connection between surfaces (2-chains) and quantum
correlations. The connection between quantum correlations and gate
operation has already been established in Theorem~1 of
\cite{RBB03}, from which the identity gate follows.

The other Clifford gates (or more precisely, CSS-gates) are derived in
a similar manner, invoking more complicated correlation surfaces.

\section{Topological considerations}
\label{TQ}

\subsection{Why can we perform non-abelian gates with surface codes?}

A limitation of the surface code \cite{Kit1, Kit2} is that only an
abelian group of gates
can be implemented fault-tolerantly by braiding operations
\cite{Kit1}. In the fault-tolerant
\QCcns, arbitrary CNOT-gates can be performed which are
non-commuting. Yet, the fault-tolerance of the \QCc is based on surface
codes. How does this fit together?

The reason why we can do non-abelian gates with a surface code in the
\QCc is that we {\em{change the topology}} of the code surface with
time. The preparation of a primal $|0\rangle$ state (dual $|+\rangle$
state) introduces a pair of primal (dual) holes into the code
surface. The corresponding measurements remove pairs of holes.

The emergence of non-abelian gates through changes in the surface
topology can be easily verified in the circuit model. Consider first the
monodromy of a primal and a dual hole as the means to entangle two
qubits of opposite type,
\begin{equation}
  \label{mono}
  \parbox[c]{5cm}{\includegraphics[width=5cm]{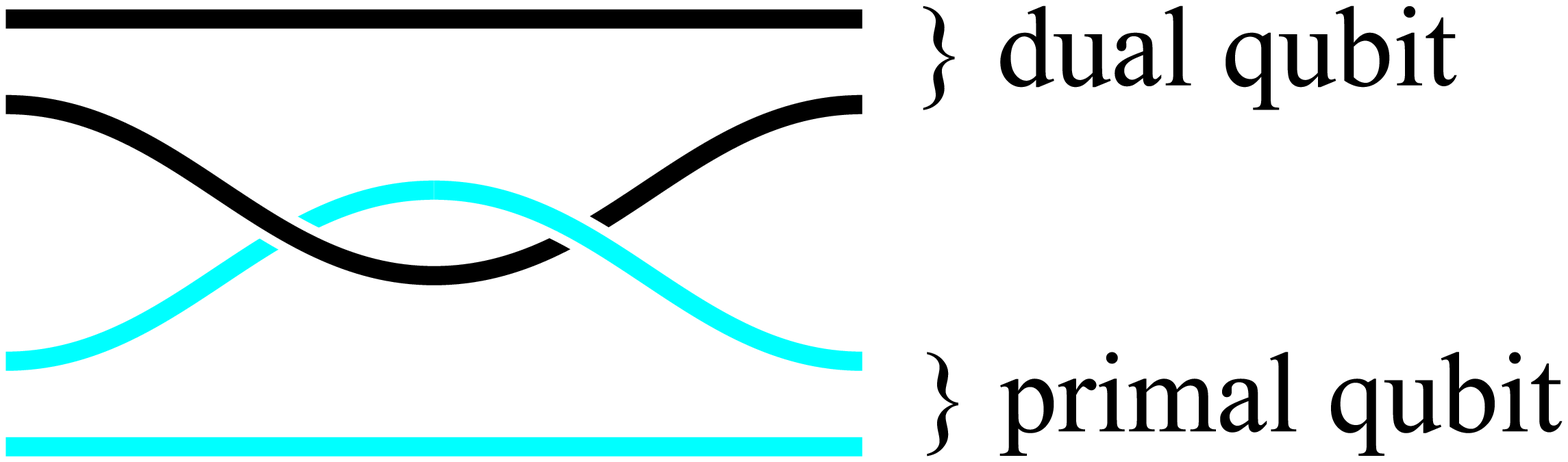}}.
\end{equation}
This operation does not change the topology of the code
surface. It can be checked with the methods described in
Section~\ref{intro} that operation  (\ref{mono}) acts as a CNOT on the
two involved qubits.  However, the
primal qubit is always the target and the dual qubit the
control. These gates are still abelian.

\begin{figure}
  \begin{center}
    \includegraphics[width=14cm]{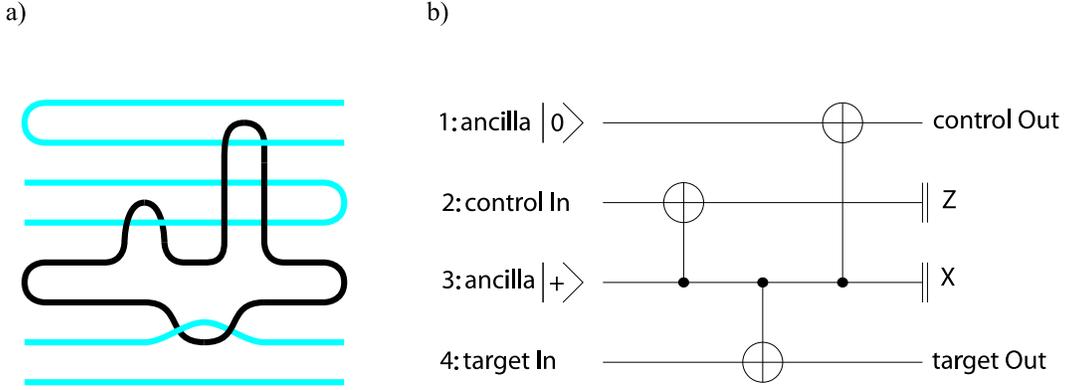}
    \caption{\label{CNOTequiv}a) Deformed version of the CNOT-gate
      displayed in Fig.~\ref{tg}a. b) Equivalent circuit, representing
      a CNOT gate between the control and target qubit.}
  \end{center}
\end{figure}

We now supplement these unitary commuting gates with non-unitary
operations, namely $X$- and $Z$-preparations and measurements. They
are obviously non-commuting and change the surface topology (See
Fig.~\ref{tg}b).
Can we construct non-commuting {\em{unitary}} operations out of this
gate set? To this end, we assemble preparations, measurements and the
monodromy operation (\ref{mono}) to the topological circuit displayed in
Fig.~\ref{CNOTequiv}a. It is a deformed version of the gate
in Fig.~\ref{tg}a. Also, it can be verified directly in the circuit
model that it represents a CNOT-gate (c.f. Fig.~\ref{CNOTequiv}b). The
direction of the CNOT can now be chosen freely, and we obtain a
non-abelian set of unitary gates.

The situation is somewhat reminiscent of the ``tilted interferometry
approach'' \cite{TI}. There, the change of a surface topology with
time is used to upgrade topological quantum computation with
Ising-anyons from non-universal to universal. In our case, the change
is from abelian to non-abelian.  As a final comment, the change of
surface topology with `time' appears as a discontinuous process. This
is an artifact of the mapping from 3 spatial dimensions to 2 spatial
dimensions plus time. In the 3D cluster picture there is no
discontinuity.

\subsection{Transforming defect configurations}

In the following we discuss equivalence transformations on the defect
configuration. Two local defect configurations are ``equivalent'' if
they have the same effect in a larger topological circuit. The
transformation rules allow us to simplify topological circuits and to
prove circuit identities.

The defects are regions in the cluster lattice
${\cal{L}}$ but for quantum information processing the details of
their shape are unimportant. Only the topology of the defect
configuration matters. As a result, the diagrams of defect strands
representing quantum gates
such as in Fig.~\ref{tg} bear a certain resemblance to link
diagrams. There are indeed similarities but there are differences,
too. The main similarity is that the line configurations representing defect
strands in these diagrams respect Reidemeister moves,
\begin{equation}
\label{Reid}
 \parbox[c]{4.5cm}{\includegraphics[width=4.5cm]{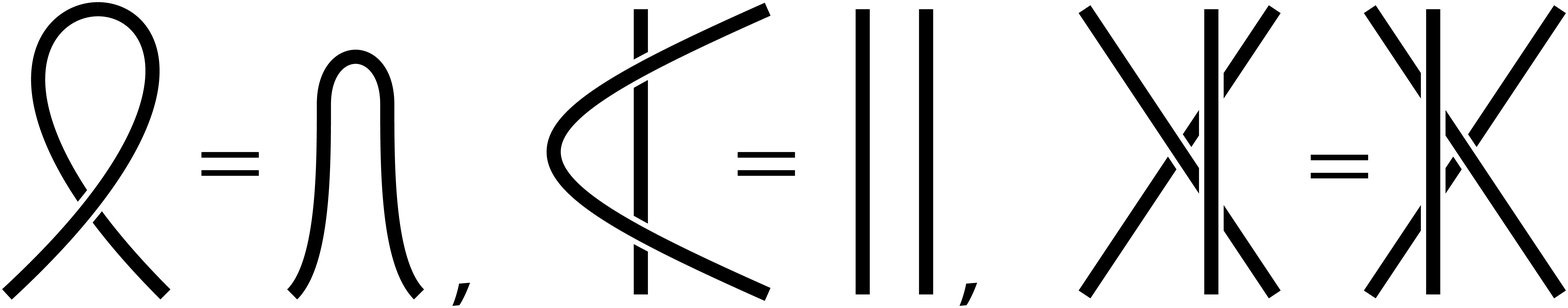}},
\end{equation}
They are valid for both types of defect and all
possible combinations. A first difference is implicit here: there
are two types of lines, primal and dual.

Next, we examine the crossings. The crossings of defect strands of
the same type are trivial,
\begin{equation}
  \label{tc}
   \parbox[c]{3cm}{\includegraphics[width=3cm]{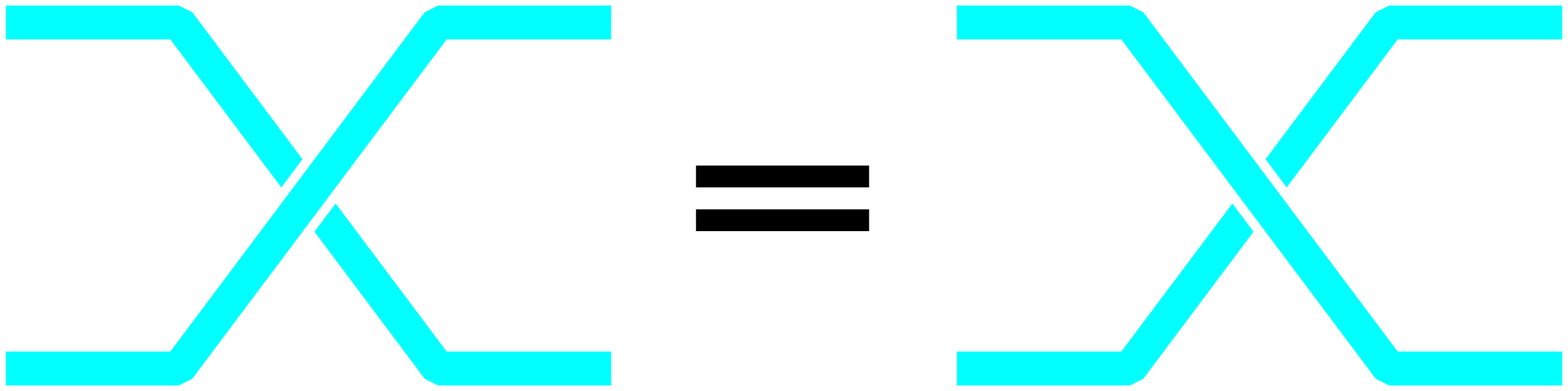}},\hspace*{3mm}
   \parbox[c]{3cm}{\includegraphics[width=3cm]{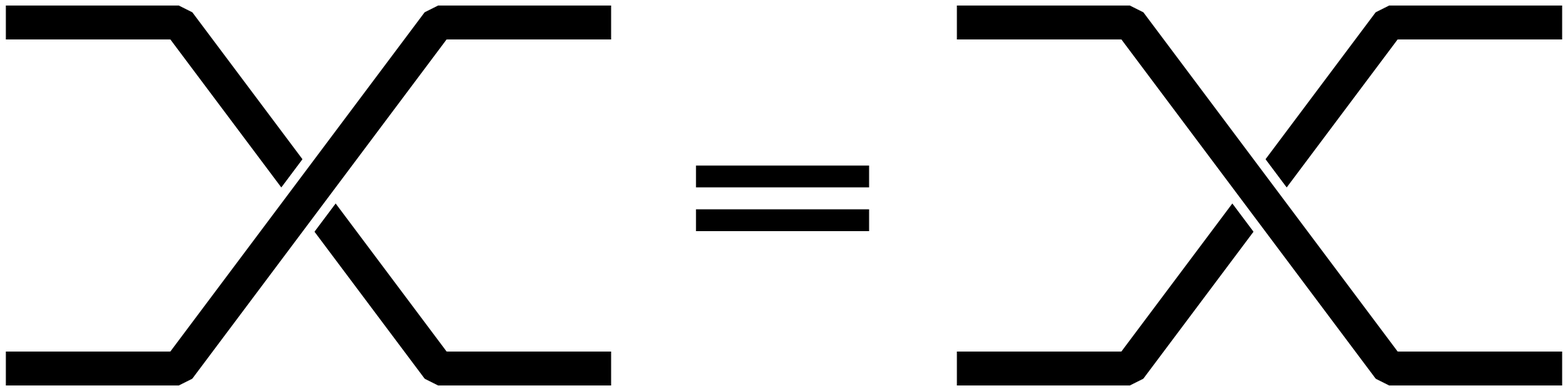}}.
\end{equation}
Only the crossing of two defects of opposite type is non-trivial; see
Eq.~(\ref{mono}). However, the double monodromy of two defect strands of
opposite type again is trivial,
\begin{equation}
  \label{doubleM}
  \parbox[c]{5cm}{\includegraphics[width=5cm]{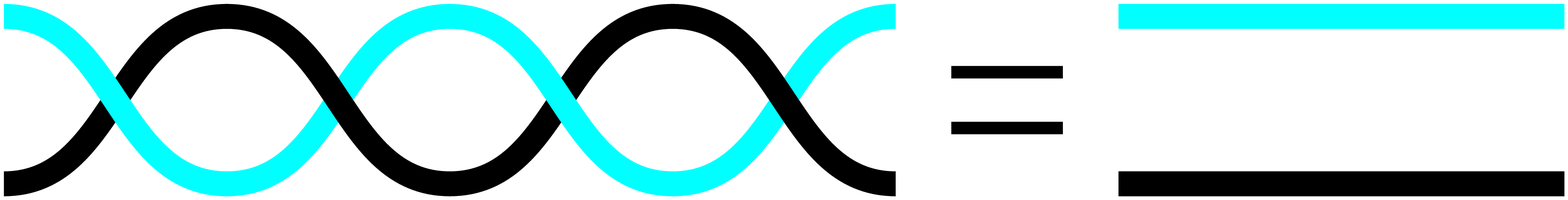}}
\end{equation}
There is a special rule for a pair of defect strands supporting a
qubit which is encircled by a defect of the opposite color. This
configuration amounts to measuring the a stabilizer generator
(\ref{SGm}) or (\ref{SGe}), respectively, of the
encoded magnetic or electric qubit. This measurement
acts as the identity operation on the code space , such that
\begin{equation}
  \label{SM}
  \parbox[c]{5cm}{\includegraphics[width=5cm]{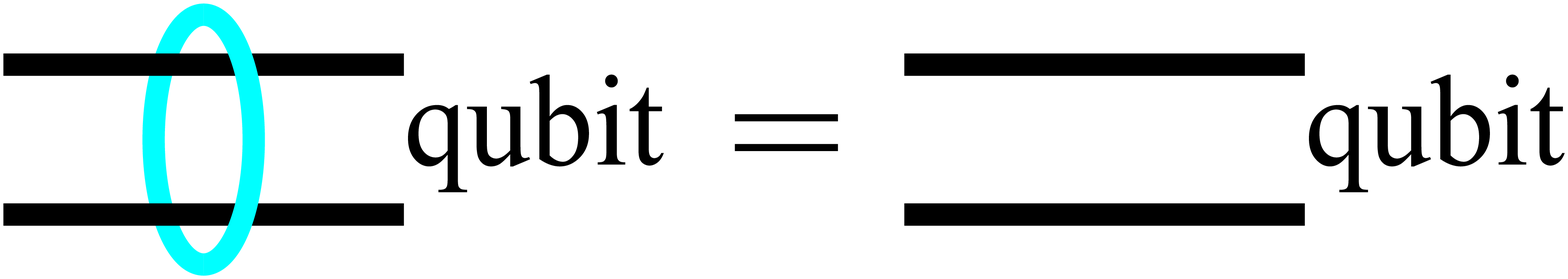}},\hspace*{3mm}
  \parbox[c]{5cm}{\includegraphics[width=5cm]{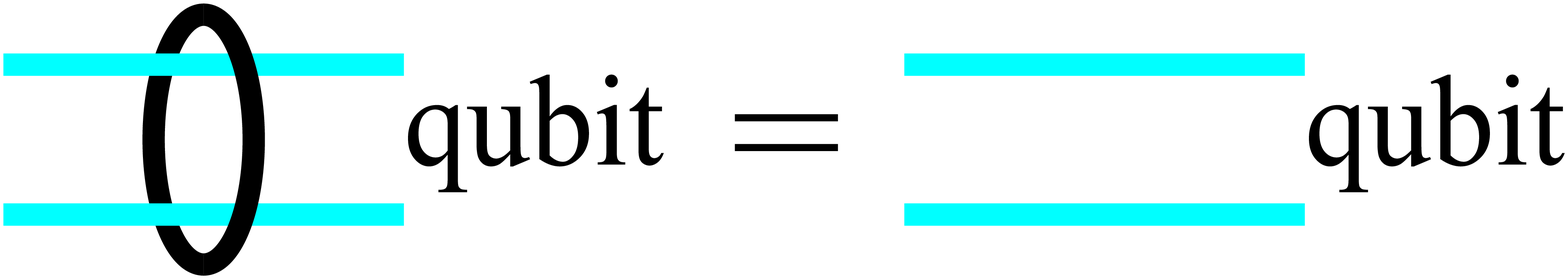}}
\end{equation}
So far, it looks as if we were discussing link diagrams with colored
components. But there is more phenomenology. Three or more defect strands
can be joined in a junction. The defect configurations thus form
graphs. Here is an equivalence transformation by
means of which junctions are introduced into the
configuration,
\begin{equation}
\label{lr}
\parbox[c]{4.5cm}{\includegraphics[width=4.5cm]{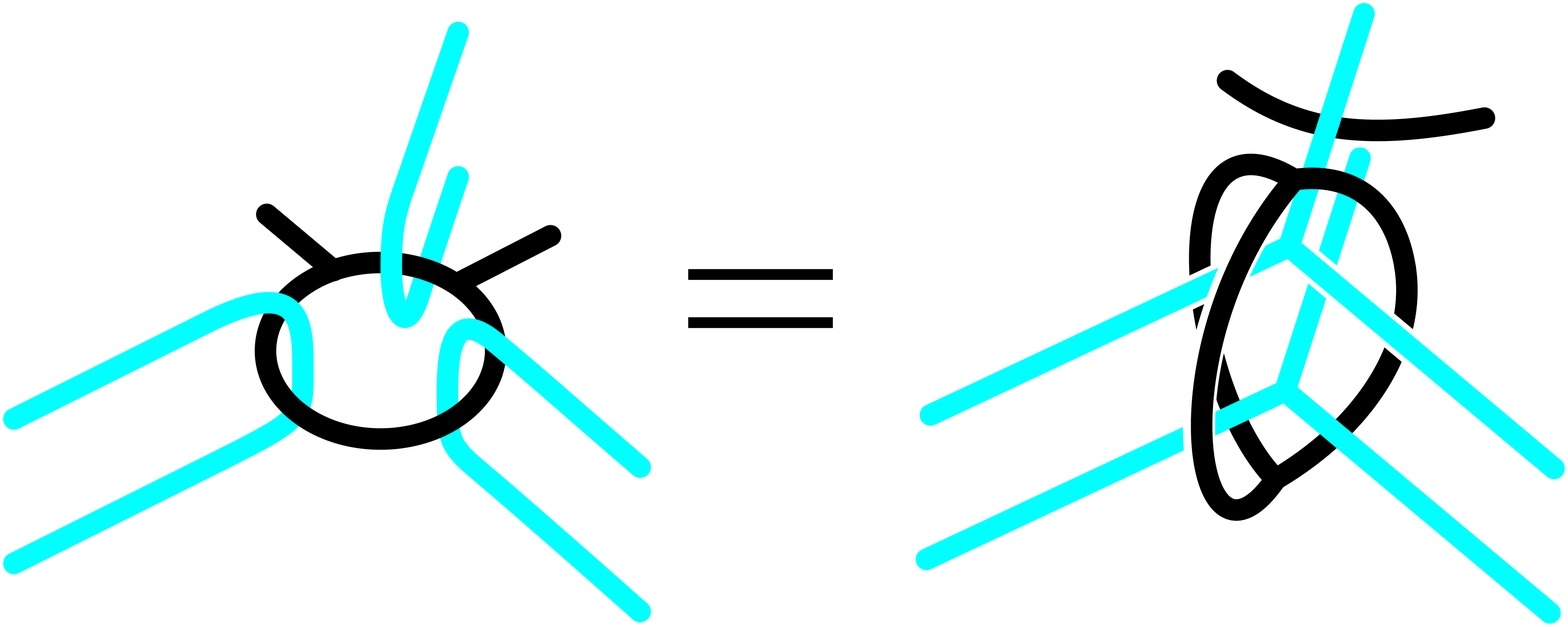}}.
\end{equation}
This is a somewhat complicated rule. The following happens here: The
dual loop on the l.h.s of (\ref{lr}) is contracted. If it
has external legs (two are shown), then these are joined in a vertex. The primal
defect strands passing the dual loop (three are shown) are cut and
reconnected. The upper and lower parts of each are joined at a
vertex. A dual cage is formed around these newly formed primal vertices.

\begin{figure}[htb]
  \begin{center}
    \includegraphics[width=5cm]{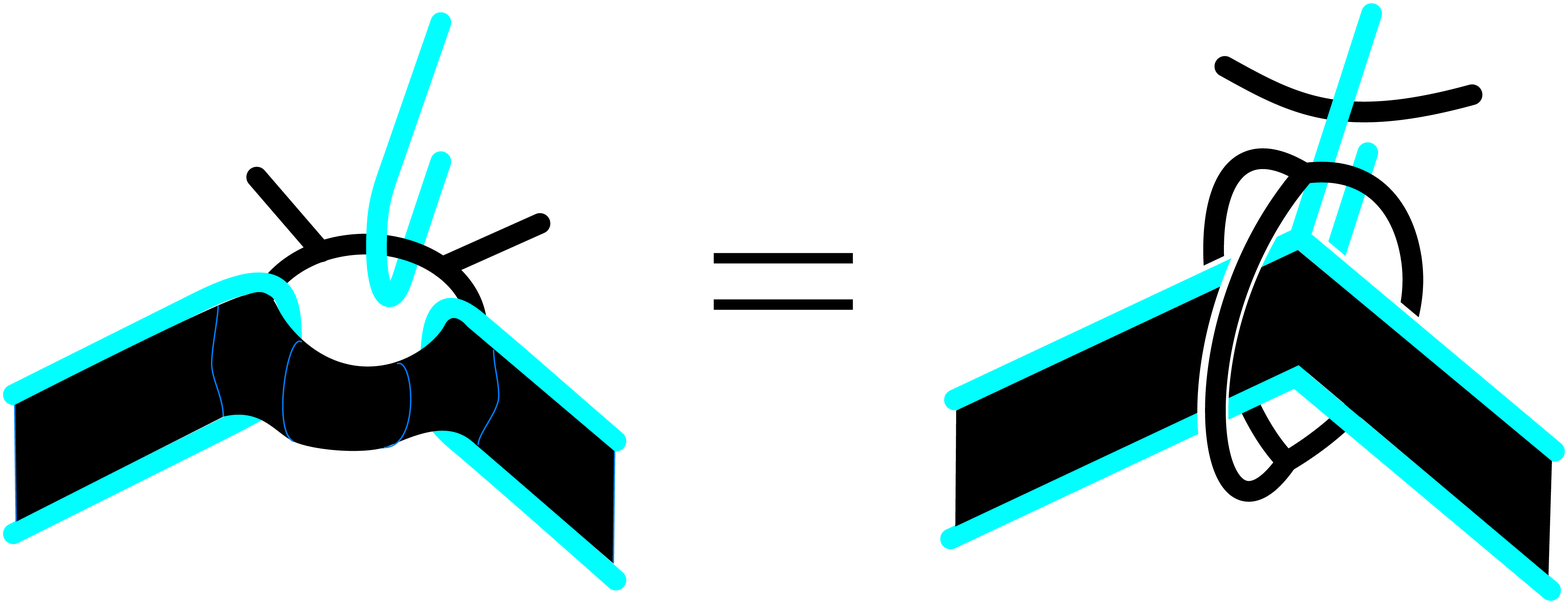}
    \caption{\label{CorrSurf}Extended correlation surface passing the
      junction. The shown
      surfaces look the same far from
      the location where surgery was performed. If the defect strands
      pairwise form qubits, the shown surface imposes a $Z\otimes
      Z$-correlation for either defect configuration.}
  \end{center}
\end{figure}

To prove that the two configurations are indeed equivalent it needs to be
checked that the set of supported correlation surfaces is the
same for each. This is beyond the scope of this paper; however, one
member of this set is displayed in Fig.~\ref{CorrSurf}. The
equivalence holds for an arbitrary number (including none) of involved primal
and dual defects. The dual
relation (primal defects $\leftrightarrow$ dual defects) also holds.

Finally, simply connected defect regions can be shrunk to a point
and removed,
\begin{equation}
  \label{point}
  \parbox[c]{4.5cm}{\includegraphics[width=4.5cm]{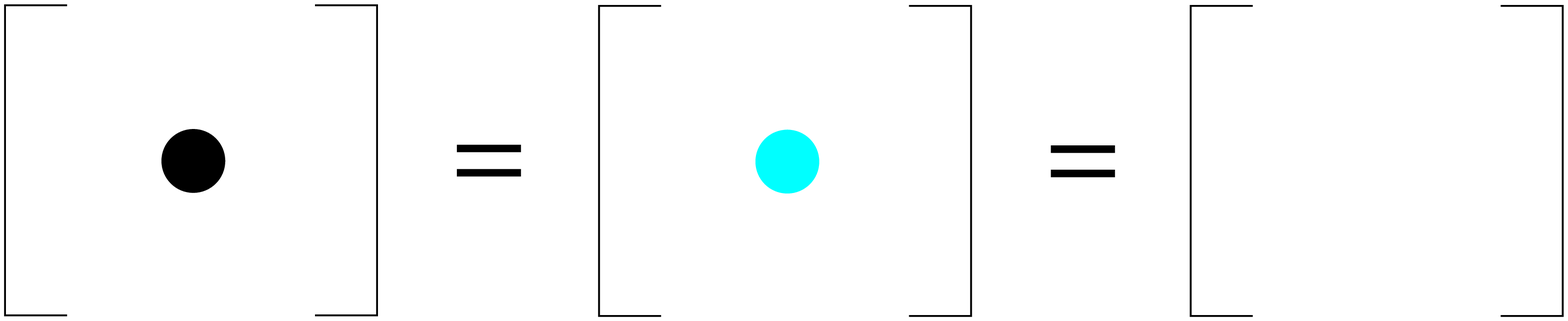}}.
\end{equation}
These rules will be used in Section~\ref{USG} to simplify sub-circuits.
To illustrate their use, we give two
examples of deriving circuit identities in a topological manner. First,
$\Lambda(X)_{c,t} |0\rangle_c \langle 0| = I_t \otimes |0\rangle_c
\langle 0|$. In the topological calculus,
\begin{center}
\begin{math}
  \parbox[c]{2.5cm}{\includegraphics[width=2.5cm]{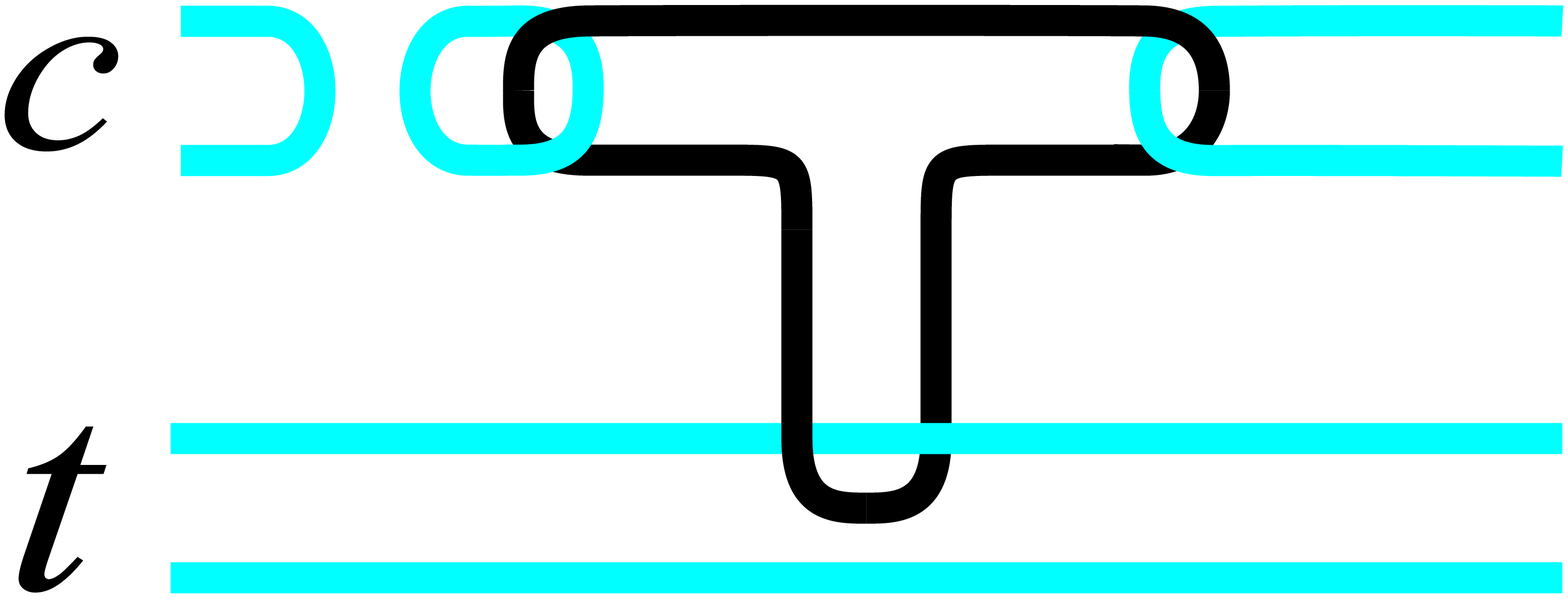}}
  \hspace{3mm} {(\ref{lr})\atop =} \hspace{3mm}
  \parbox[c]{2.5cm}{\includegraphics[width=2.5cm]{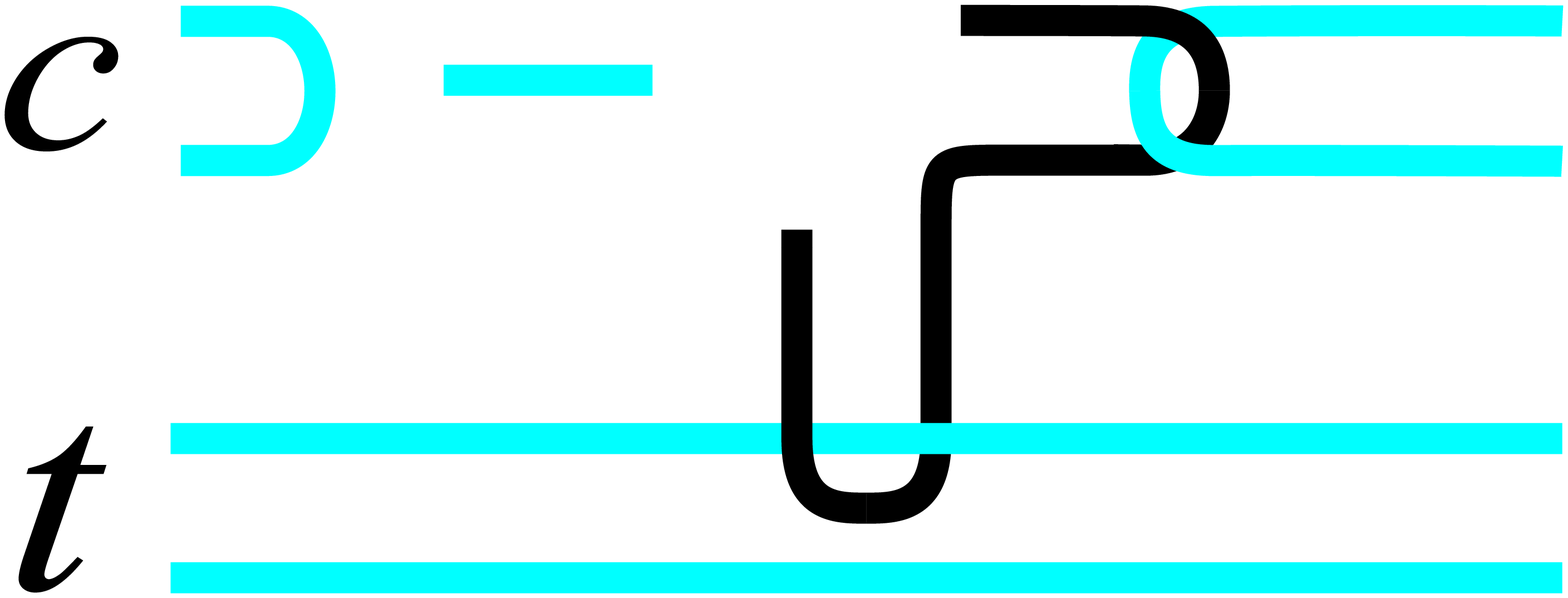}}
  \hspace{3mm}  {(\ref{point})\atop =} \hspace{3mm}
  \parbox[c]{2.5cm}{\includegraphics[width=2.5cm]{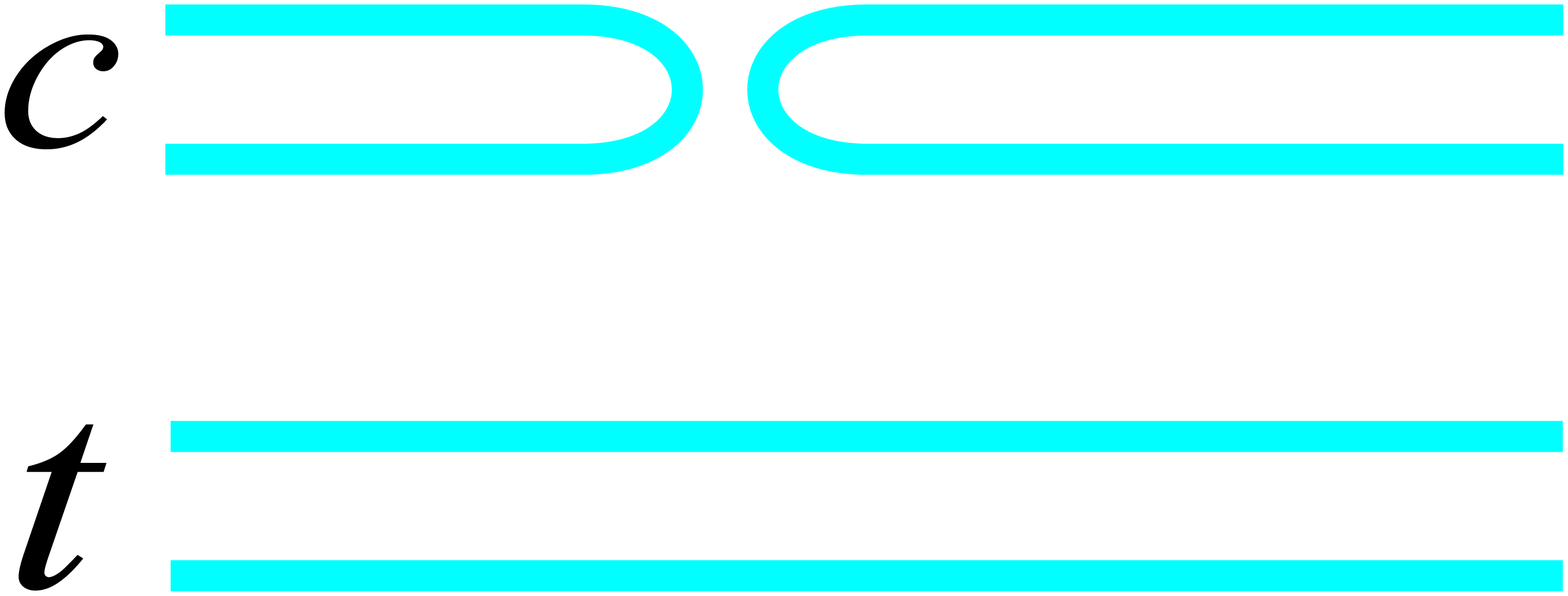}}
\end{math}
\end{center}
Second, $\Lambda(X)_{a,b} \Lambda(X)_{b,a} \Lambda(X)_{a,b} =
\mbox{SWAP}(a,b)$. In the topological calculus,
\begin{center}
\begin{math}
  \begin{array}{cccc}
  \parbox[c]{5.4cm}{\includegraphics[width=5.4cm]{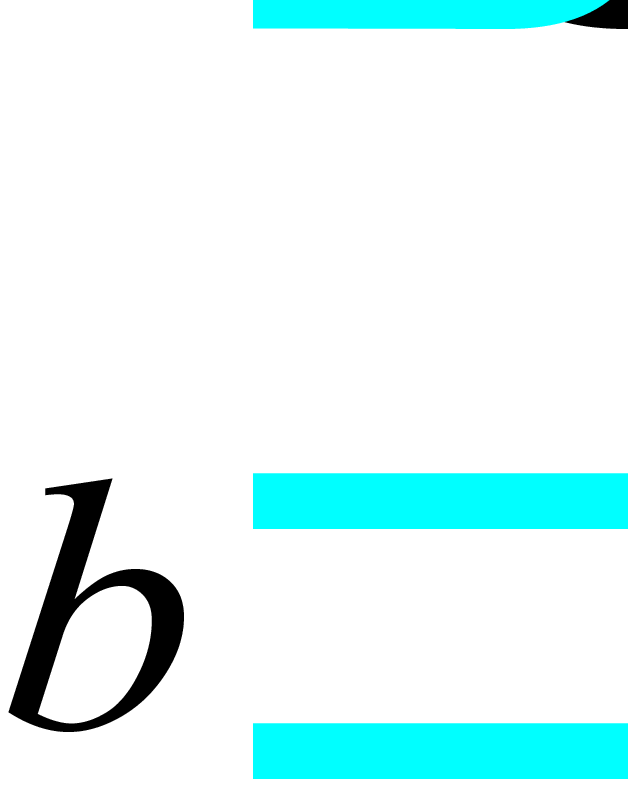}}
  &{(\ref{lr})\atop =}&
   \parbox[c]{5.4cm}{\includegraphics[width=5.4cm]{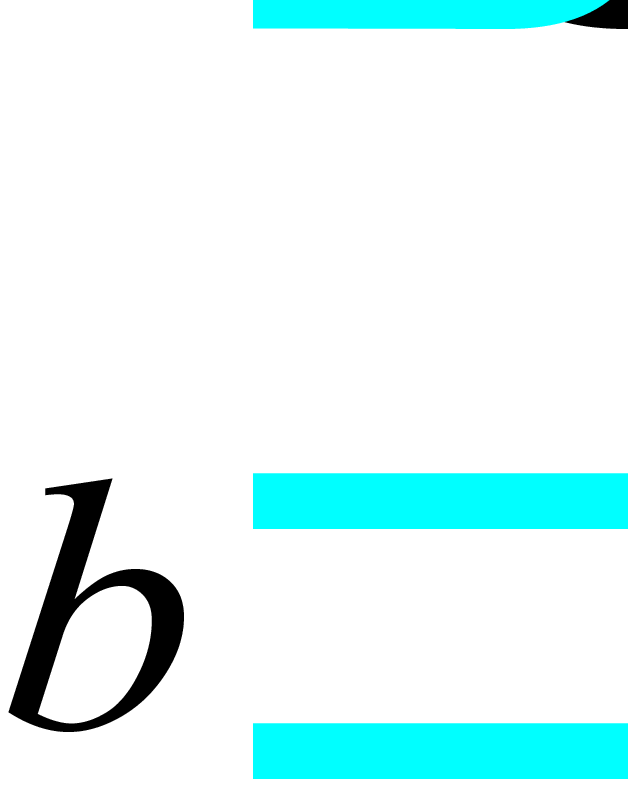}}
  & {(\ref{tc})\atop =}\\
   \parbox[c]{5.4cm}{\includegraphics[width=5.4cm]{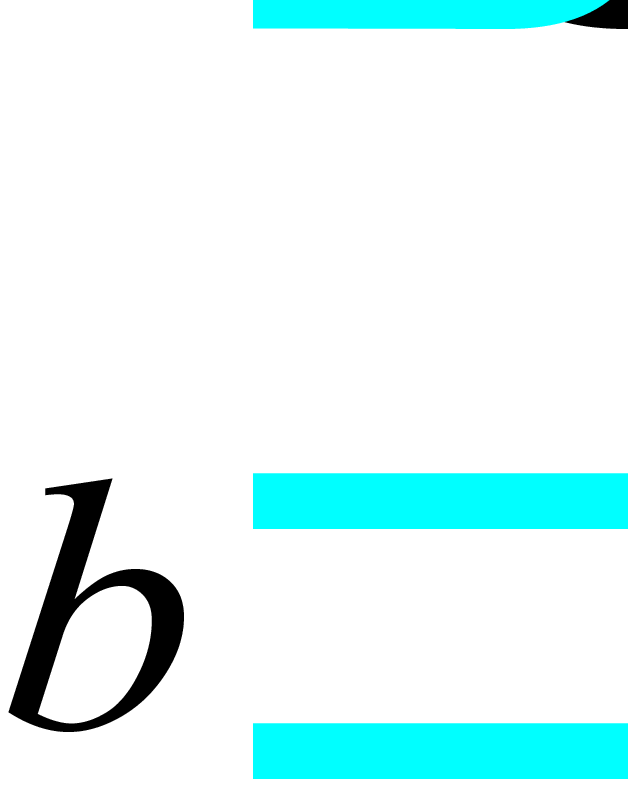}}
  &{(\ref{lr})\atop =}&
  \parbox[c]{5.4cm}{\includegraphics[width=5.4cm]{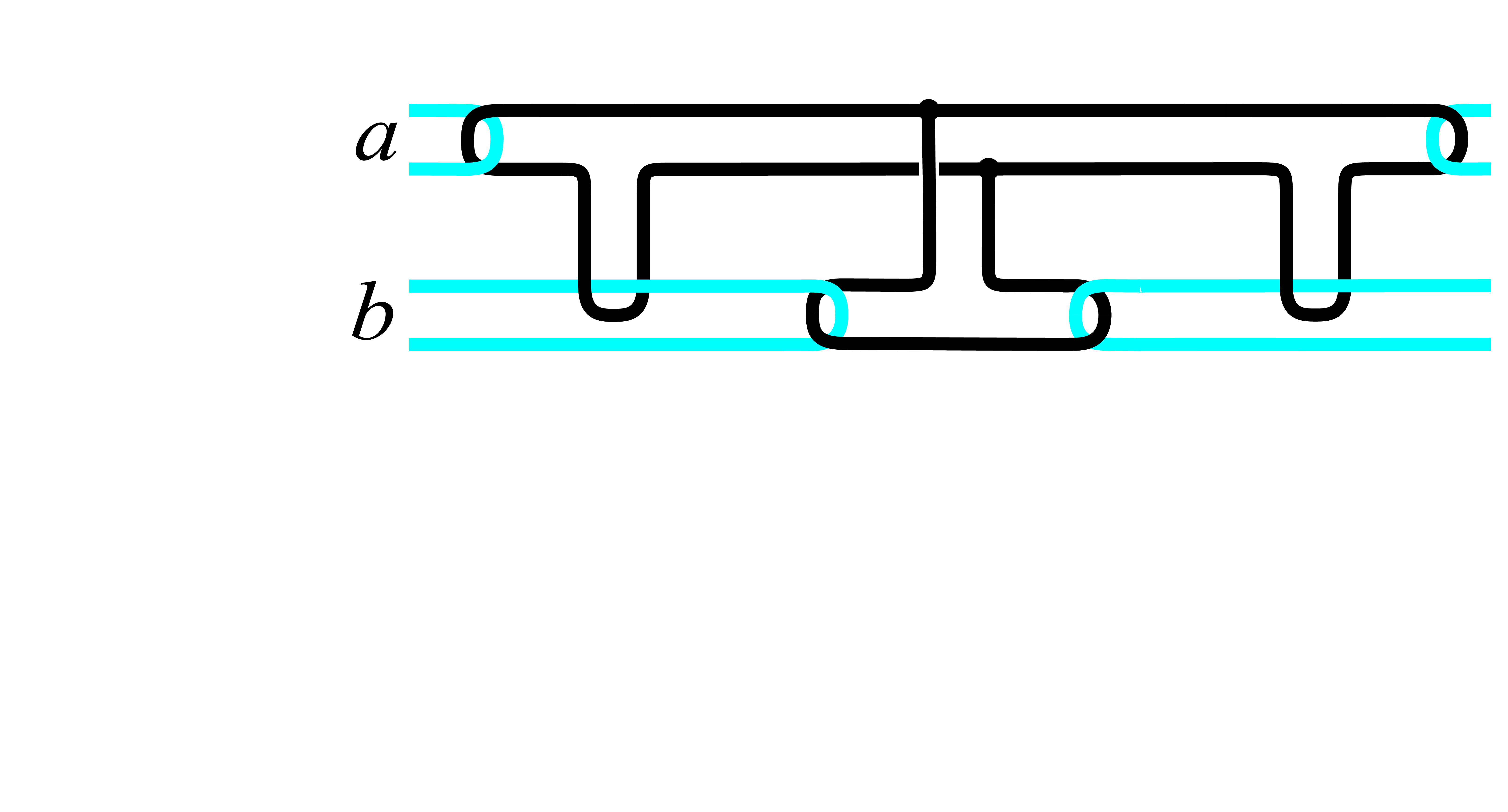}}
  &{(\ref{tc})\atop =}\\
   \parbox[c]{5.4cm}{\includegraphics[width=5.4cm]{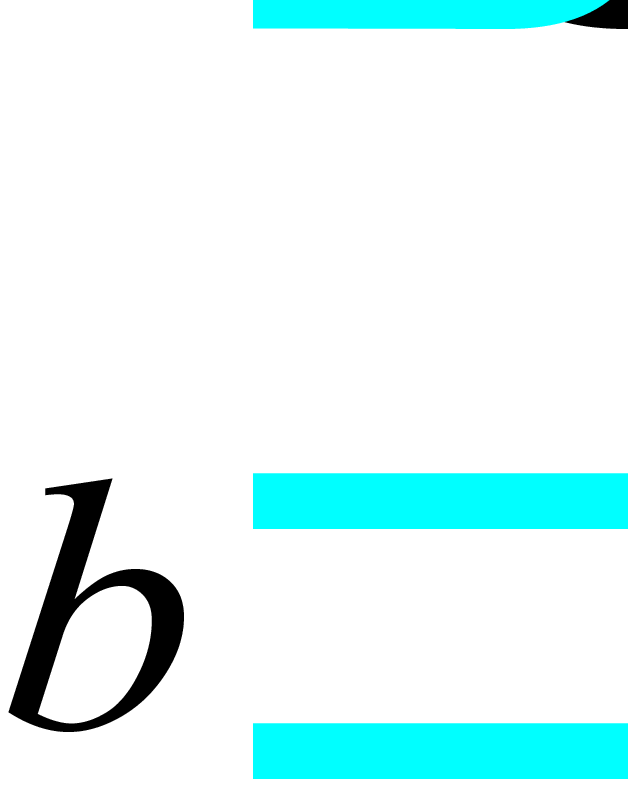}}
  &{(\ref{doubleM})\atop =}&
  \parbox[c]{5.4cm}{\includegraphics[width=5.4cm]{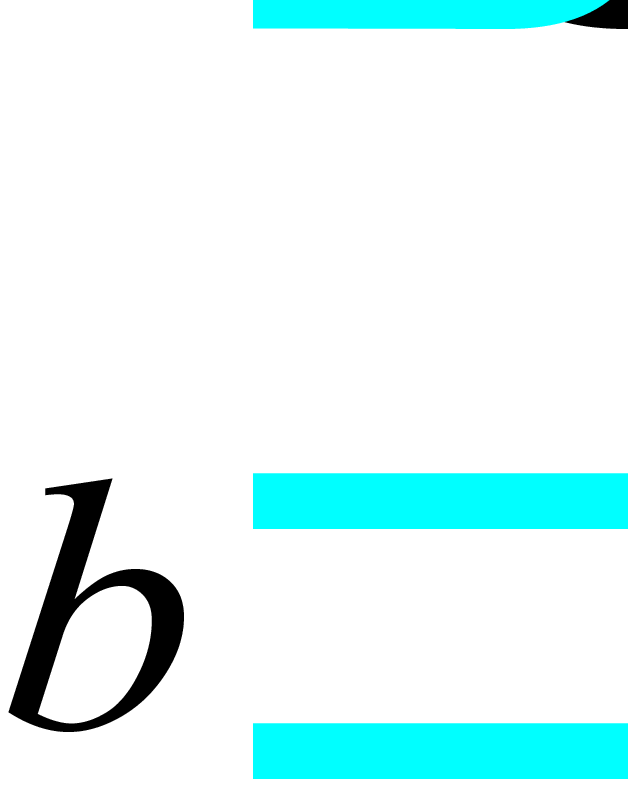}}
  &{(\ref{doubleM}) \atop =}\\
   \parbox[c]{5.4cm}{\includegraphics[width=5.4cm]{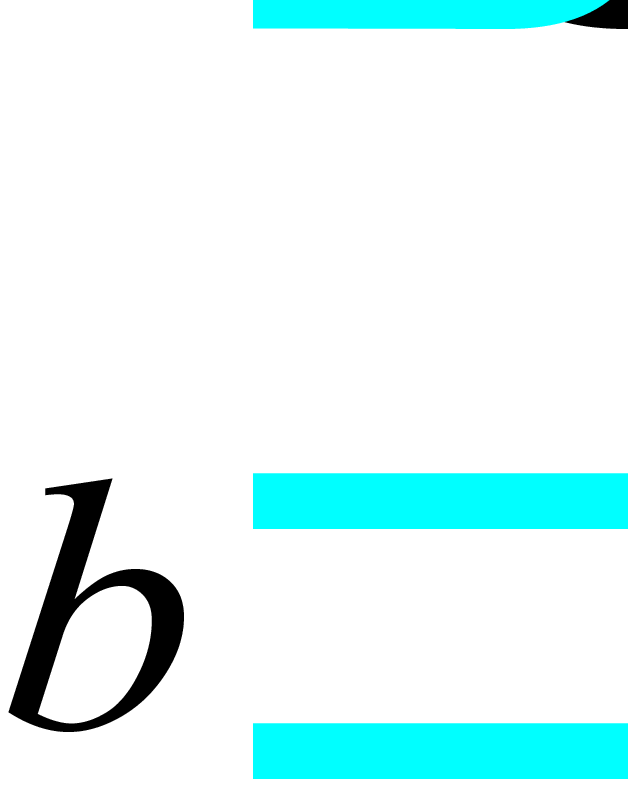}}
  &{(\ref{lr}) \atop =}&
  \parbox[c]{5.4cm}{\includegraphics[width=5.4cm]{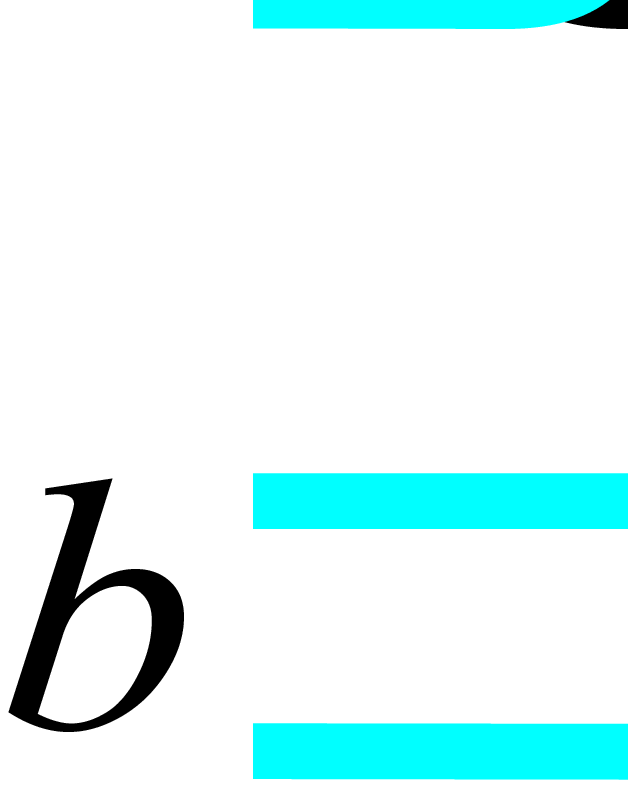}}
  &{(\ref{Reid}) \atop =}\\
   \parbox[c]{5.4cm}{\includegraphics[width=5.4cm]{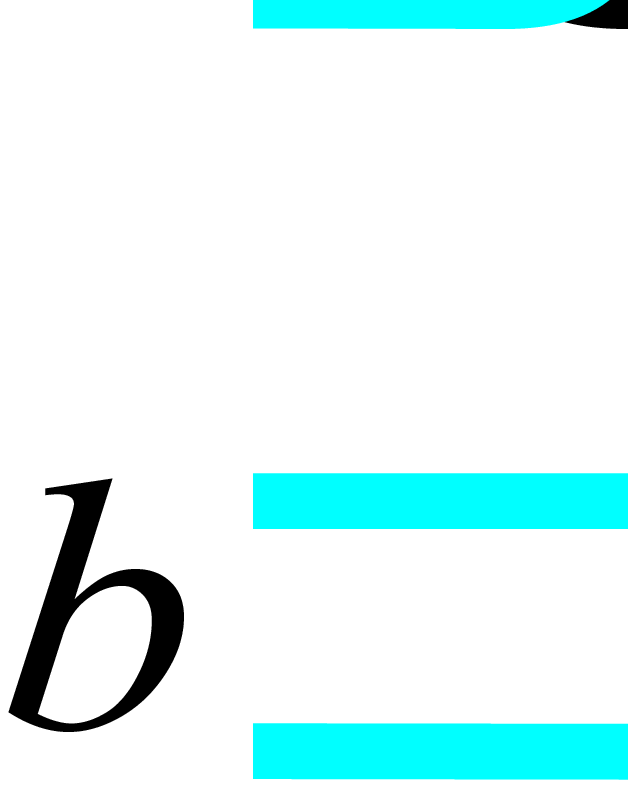}}
  &{(\ref{doubleM})\atop =}&
  \parbox[c]{5.4cm}{\includegraphics[width=5.4cm]{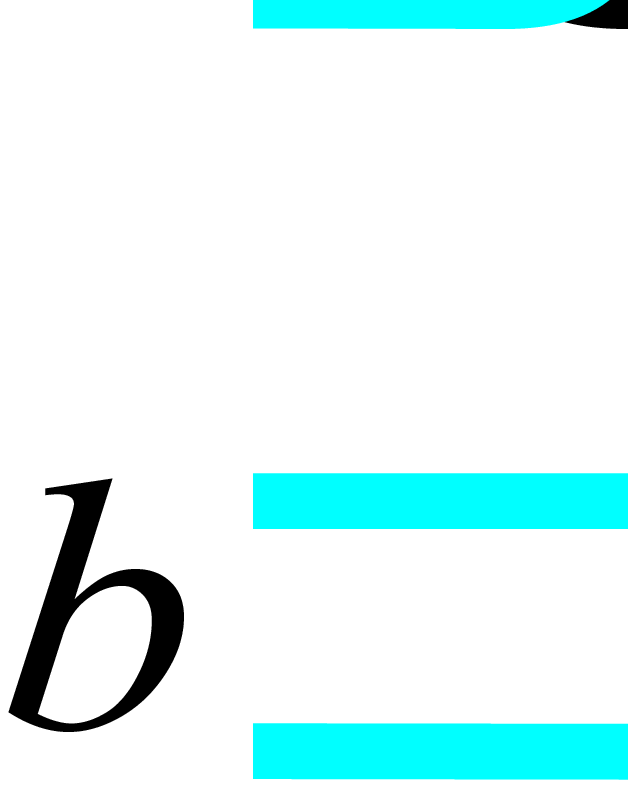}}
  &{(\ref{lr})\atop =}\\
   \parbox[c]{5.4cm}{\includegraphics[width=5.4cm]{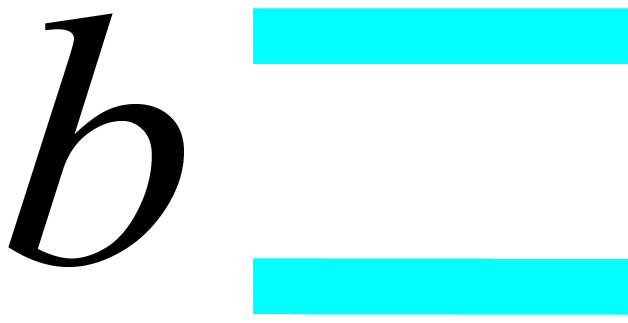}}
  &{(\ref{tc},\ref{SM})\atop =}&
  \parbox[c]{5.4cm}{\includegraphics[width=5.4cm]{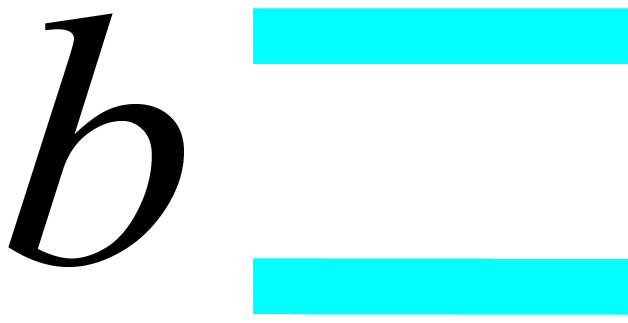}}.
  &\\
  \end{array}
\end{math}
\end{center}

\section{Completing the universal set of gates}
\label{USG}

The topologically protected gates, the CNOT and
preparation/measurement in the $X$- and $Z$-eigenbasis, are shown in
Fig.~\ref{tg}. The $X$- and $Z$-measurements are obtained by reversing
the time-arrow in the corresponding state preparations.

\begin{figure}[b]
  \begin{center}
    \includegraphics[width=4cm]{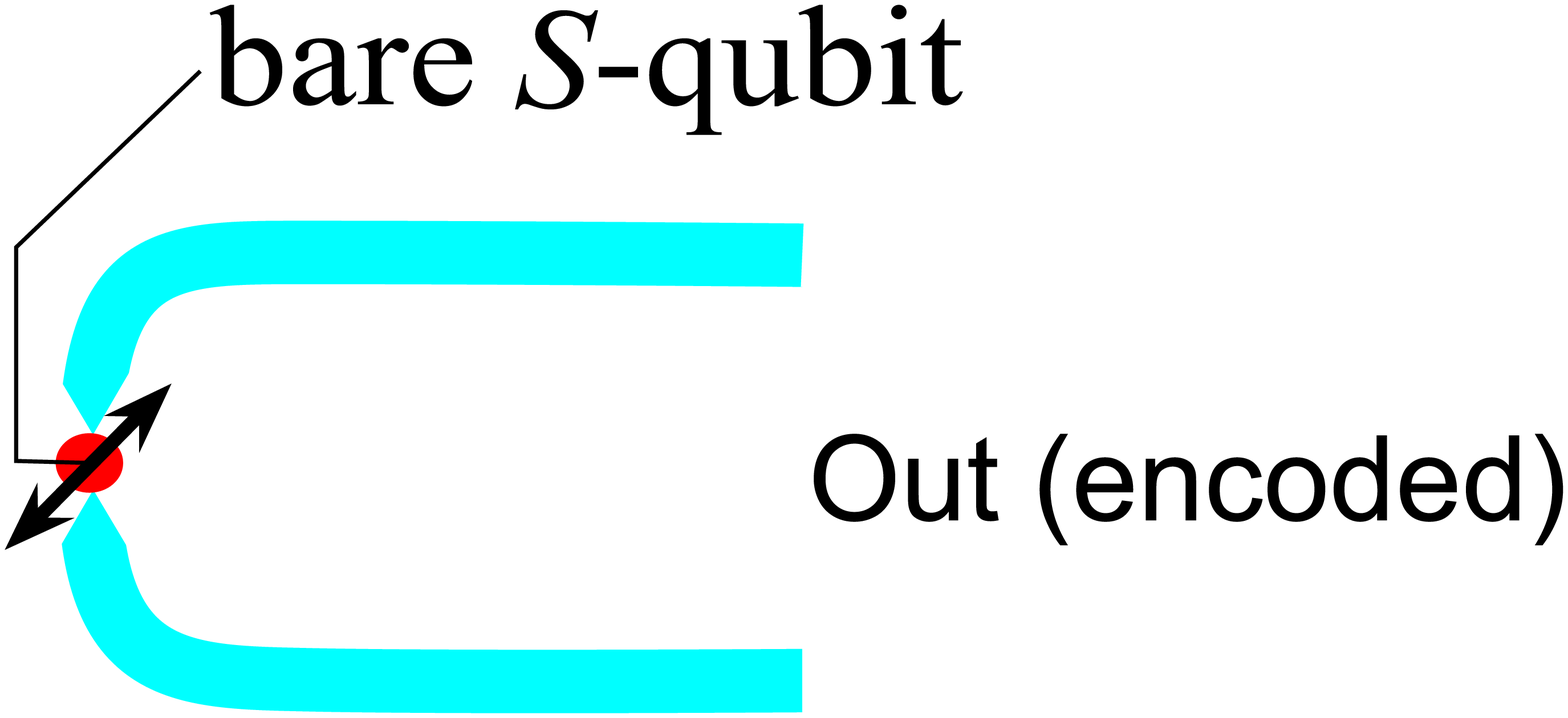}
    \caption{\label{AP}Preparation of the ancillas $|Y\rangle$ and
      $|A\rangle$, encoded with the surface code. To obtain
      $|Y\rangle$ and $|A\rangle$, the singular qubit $S$ is measured
      in the eigenbasis of $Y$ or $(X+Y)/\sqrt{2}$, respectively.}
  \end{center}
\end{figure}

We can complete these operations to an universal set by adding
$\exp(i\frac{\pi}{8}Z)$, $\exp(i\frac{\pi}{4}Z)$, $\exp(i\frac{\pi}{4}X)$. The
fault-tolerant realization of these gates requires error-free ancilla states
$|Y\rangle:= (|0\rangle + i|1\rangle)/\sqrt{2}$ and $|A\rangle:=(|0\rangle +
e^{i\pi/4} |1\rangle)/\sqrt{2}$. These states are first created in a
noisy fashion using the element displayed in Fig.~\ref{AP}, and then
distilled \cite{BK04}. For details, see
Section \ref{OH} and Appendix \ref{RM}.

\begin{figure}[htb]
  \begin{center}
    \includegraphics[width=9cm]{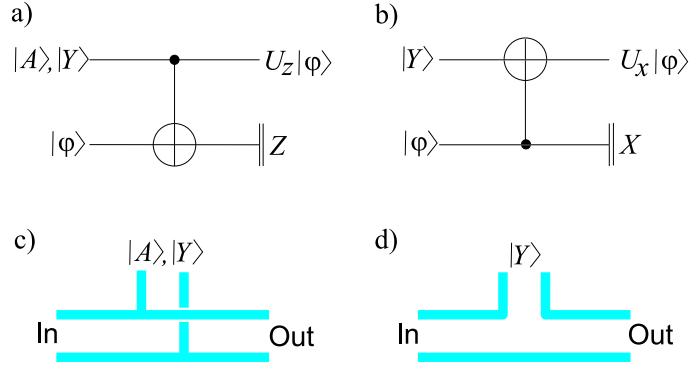}
    \caption{\label{gadgets}One-qubit rotations. a) Circuit for
      performing $U_Z$-gates using the ancillas $|Y\rangle$,
      $|A\rangle$.  b) Circuit for
      performing $U_X$-gates using the ancilla $|Y\rangle$. c) and d)
      show the corresponding defect configurations.}
  \end{center}
\end{figure}

Once the ancilla states $|A\rangle$ and $|Y\rangle$ have been distilled
they are used in the circuits of Fig.~\ref{gadgets}a,b to produce the desired
gates. The gate $\exp(i\frac{\pi}{8} Z)$ is probabilistic and succeeds with
probability 1/2. Upon failure, the gate $\exp(-i\frac{\pi}{8} Z)$ is
applied instead, which can be corrected for by a subsequent operation
$\exp(i\frac{\pi}{4} Z)$. The latter gate is
deterministic modulo Pauli operators, which suffices for the \QCcns.

Their fault-tolerant \QCcns-realizations for the above gates are shown
in Fig.~\ref{gadgets}c,d. These realizations are obtained from pasting the
standard elements for the CNOT and measurement together and subsequently
applying the defect transformation rules (\ref{Reid}) - (\ref{point}).

\section{Mapping to a two-dimensional system}
\label{2D}

The dimensionality of the
spatial layout can be reduced by one if the cluster is created slice
by slice. That is, we convert
the `simulated time'-axis---introduced as a means to explain the
connection with surface codes---into real time.
Under this mapping, cluster qubits located on time-like (space-like) edges of
${\cal{L}}$, $\overline{\cal{L}}$ become syndrome qubits (code qubits)
which are (are not) periodically measured.

\begin{figure}
  \begin{center}
    \begin{tabular}{ll}
      a) & b)\\
      \parbox[c]{5cm}{\includegraphics[height=5cm]{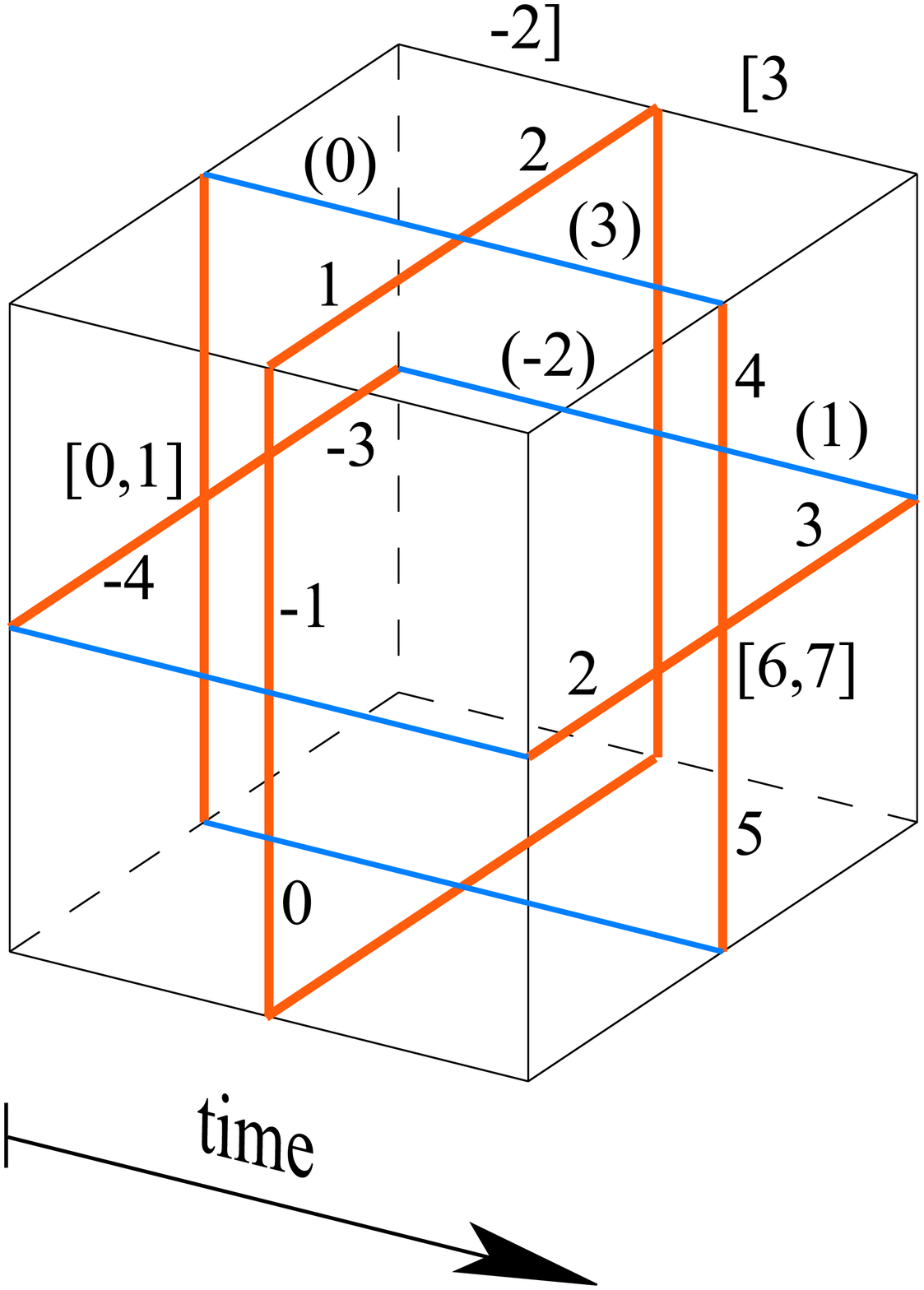}}&
      \parbox[c]{6cm}{\includegraphics[height=4cm]{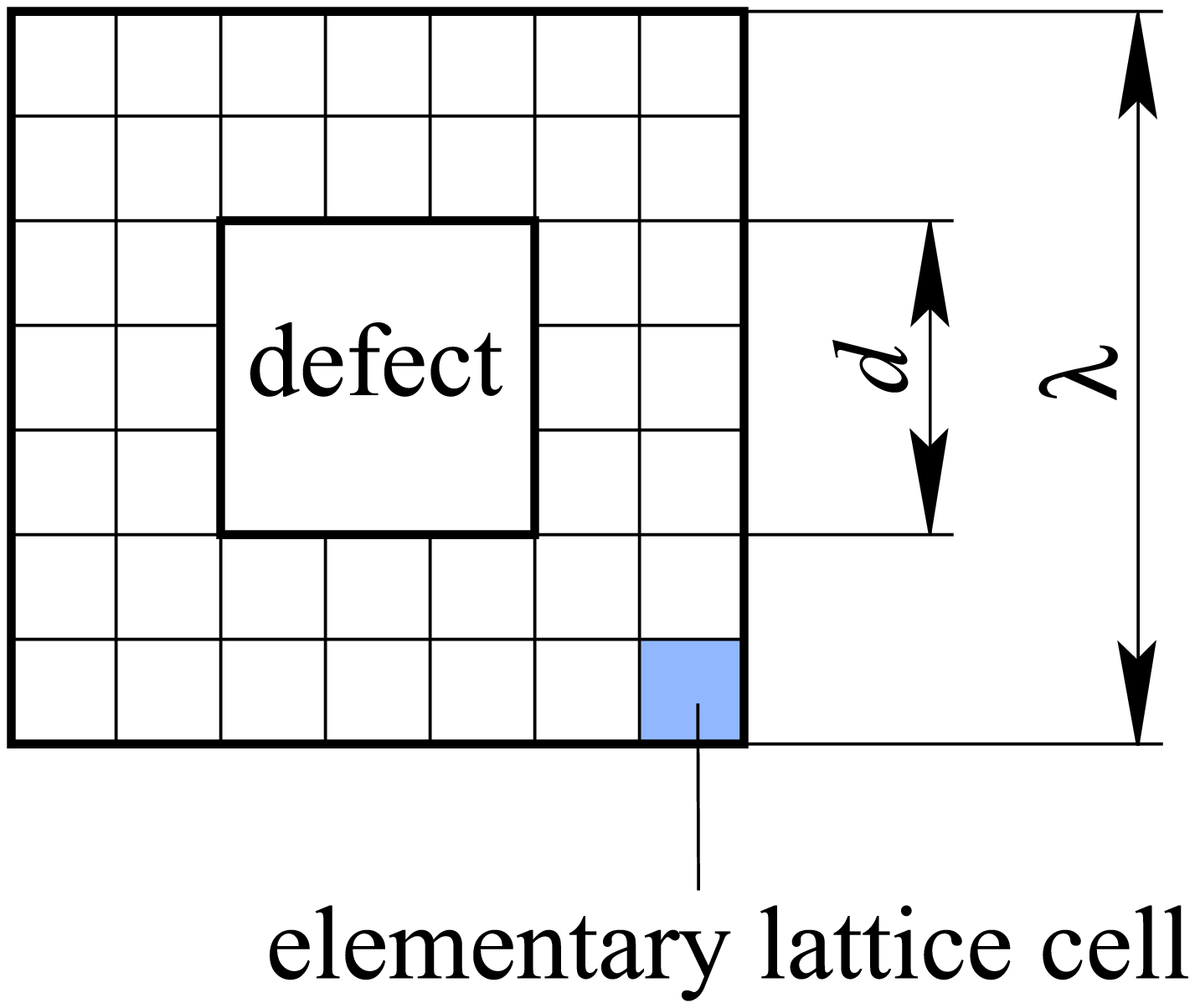}}
    \end{tabular}
    \caption{\label{Cell2D}a) Temporal order of operations in $V$ after
      the mapping to 2D. Shown is the elementary cell of the 3D
      lattice with one axis converted to time.  The labels on the edges
      denote the time steps at which
      the corresponding $\Lambda(Z)$-gate is performed. The labels at
      the syndrome vertices denote measurement and
      (re-)preparation times  $[t_M, t_P]$, and the labels  $(t_H)$
      denote times for Hadamard gates. The pattern is periodic
      in space, and in time with period six. b) The logical cell. It is
      rescaled from
      the elementary cell of the lattice ${\cal{L}}$ by a factor of
      $\lambda$ in each direction. The defects have cross-sections
      $d\times d$.}
  \end{center}
\end{figure}

Most important is the region $V$ in which we have topological error
protection. Therein, space-like oriented $\Lambda(Z)$-gates
remain and time-like oriented  $\Lambda(Z)$ gates are mapped into
Hadamard gates. The temporal order of operations is displayed in
Fig.~\ref{Cell2D}. Note that every qubit is acted upon by an operation in every
time step. The mapping to the two-dimensional structure has no
impact on the information processing. In particular, the
error correction procedure is still the same as in fault-tolerant
quantum memory with the toric code.

In the 3D version, we use $|+\rangle$-preparations and
$\Lambda(Z)$-gates for the creation of $|\phi\rangle_{\cal{L}}$, and
subsequently perform local $X$, $X\pm Y$, $Y$ and $Z$-measurements. We
now give the complete mapping for these operations to the 2+1
dimensional model.

{\em{1. Space-like edges (primal and dual).}} We group together the
respective $|+\rangle$-preparation, measurement and trailing
time-like oriented $\Lambda(Z)$-gate, and denote the combination by
$\{|+\rangle, \Lambda(Z), P\}$. If the measurement on the trailing end
of $\Lambda(Z)$ is in the $Z$-basis, then
\begin{equation}
  \{|+\rangle, \Lambda(Z), P\} \longrightarrow P.
\end{equation}
Otherwise,
\begin{equation}
  \begin{array}{lcl}
    \{|+\rangle, \Lambda(Z), P_X\} & \longrightarrow & H,\\
    \{|+\rangle, \Lambda(Z), P_{X\pm Y}\} & \longrightarrow & H
    e^{i\frac{\pi}{8} Z},\\
    \{|+\rangle, \Lambda(Z), P_Y\} & \longrightarrow & H
    e^{i\frac{\pi}{4} Z},\\
    \{|+\rangle, \Lambda(Z), P_Z\} & \longrightarrow & P_X.
  \end{array}
\end{equation}

{\em{2. Time-like edges (primal and dual).}} For each such edge, we
group together the respective preparation and measurement, and denote
the combination by $\{|+\rangle, P\}$. Then,
\begin{equation}
  \label{TLE}
  \begin{array}{lcll}
    \{|+\rangle, P_Z\} & \longrightarrow & I,\\
    \{|+\rangle, P\} & \longrightarrow & \{|+\rangle,P\}, &\mbox{for
    } P\neq P_Z.
  \end{array}
\end{equation}

{\em{3. Space-like oriented $\Lambda(Z)$-gates.}}
\begin{equation}
\label{slcPh}
 \begin{array}{lcl}
   \Lambda(Z)_{a,b} & \longrightarrow & \Lambda(Z)_{a,b}.
  \end{array}
\end{equation}

{\em{Remark 1.}} No qubit in the scheme is ever idle between
preparation and measurement. The identity in the first line of
Eq. (\ref{TLE}) can be replaced by the 1-qubit completely depolarizing map
without affecting the scheme. The respective qubit will be
re-initialized before its next use.

{\em{Remark 2.}} From the perspective of information processing, the
space-like oriented gates $\Lambda(Z)_{a,b}$
in (\ref{slcPh}) have no effect if $a \in D\, \vee \, b \in D$. They may
consequently be left out. Keeping these redundant gates in the scheme,
however, does not affect the threshold; see remark 4.
We keep the redundant $\Lambda(Z)$-gates in order to {\em{maintain
    translational invariance}} of the (Ising) qubit-qubit interaction.

{\em{Remark 3.}} For physical realization of the scheme with cold
atoms in an optical lattice it may be preferable to use a double-layer 2D
structure instead of a single layer. The advantage then is that all
qubits within one layer, including the $S$-qubits, {\em{can be read
    out simultaneously}}. One `clock cycle' consists of the following
steps: 1) Ising interaction/ $\Lambda(Z)$-gates between all pairs of
nearest neighboring qubits in the lattice; 2) Simultaneous measurement
of all qubits in layer $a$, re-preparation of all qubits in layer $b$;
3) Same as 1); 4) Same as 2), with $a \leftrightarrow b$.

Ideally, one would use a bcc lattice half a
cell thick but an sc lattice one cell thick also works. In the latter
case, some redundant $\Lambda(Z)$gates/ Ising-type interactions and
$Z$-measurements increase the number of error sources and thus moderately
reduce the error threshold.

\section{Fault-tolerance and threshold}
\label{FTT}

\paragraph{Error model.} We assume the following:
\begin{enumerate}
\item{Erroneous operations are
modeled by perfect operations preceded/followed by a partially depolarizing
single- or two-qubit error channel
\begin{eqnarray}
T_1 &=& (1-p_1)[I]+p_1/3\,([X]+[Y]+[Z]), \nonumber\\
T_2 &=& (1-p_2)[I]+p_2/15\,([X_aX_b]+..+[Z_aZ_b]).\nonumber
\end{eqnarray}}
\item{The error
sources are a) faulty preparation of the individual qubit states $|+\rangle$
(error probability $p_P$), b) erroneous Hadamard-gates (error
probability $p_1$), c) erroneous $\Lambda(Z)$-gates (error
probability $p_2$),
and d) imperfect measurement (error probability $p_M$).}
\item{Classical processing is instantaneous.}
\end{enumerate}
When calculating a threshold, we consider
all error sources to be equally strong, $p_1=p_2=p_M=p_P:=p$, such that
the noise strength is described by a single parameter $p$. Storage
errors need  not be considered because no qubit is ever idle between
preparation and measurement.

\paragraph{Error correction.} The three relevant facts about
fault-tolerance in the \QCc are
\begin{itemize}
  \item{The error correction in $V$ is topological. It can be mapped
      to the {\em{random plaquette $\mathbb{Z}_2$-gauge model}} (RPGM)
      in three dimensions \cite{DKLP}. If there are
      non-trivial error cycles of finite
      smallest length $l$ then, below the error threshold, the
      probability of error $\epsilon_{top}$ is
      \begin{equation}
        \label{expsup}
        \epsilon_{top} \sim \exp(-\kappa(p) l).
      \end{equation}}
  \item{The topological error correction breaks down near the singular
  qubits. This results in an effective error on the $S$-qubits that
  needs to be taken care of by an additional correction method. This
  effective error is {\em{local}} because the $S$-qubits are far
  separated from another \cite{RHG}.}
\item{The cluster region $D$ need not be present at all. It is
      initially included to keep the creation procedure of the cluster
      state translation invariant, and subsequently removed by local
      $Z$-measurement of all qubits in $D$. The purpose of $D$ is to
      create non-trivial boundary conditions for the remaining cluster.}
\end{itemize}
The fault-tolerance threshold associated with the RPGM is about
$3.2\times 10^{-2}$ \cite{Japs}, for a strictly local error model with
one source. Also see \cite{J2}. The threshold estimates given in this
paper are based on the {\em{minimum weight chain matching algorithm}}
\cite{AlgoEff} for error correction. This algorithm yields a slightly
smaller threshold of 2.9\% \cite{Harri} but is compuationally efficient.

\begin{figure}
  \begin{center}
    \includegraphics[width=9cm]{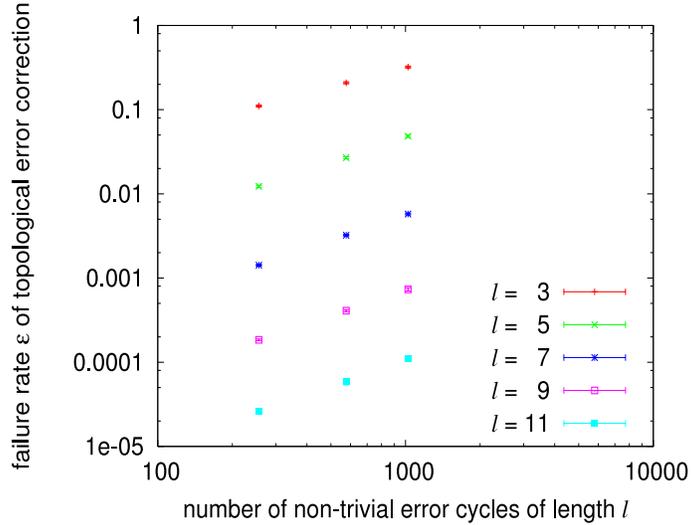}
    \caption{\label{ExpDec}Exponential decay of the failure rate
    $\epsilon$ of the topological error correction 
    as a function of the length $l$ of
    the shortest non-trivial error
    cycle and the number $N$ of such cycles, at $p = 1/3\, p_c$. 
    }
  \end{center}
\end{figure}

Concerning the exponential decay of error probability in
$V$ in Eq.~(\ref{expsup}), the dominant behavior is both predicted
from a Taylor expansion of $\epsilon_{top}$ in terms of the physical
error rate $p$ (truncated at lowest contributing order \cite{RHG})
and confirmed by numerical simulation (see Fig.~\ref{ExpDec}).
Beyond the dominant exponential decay there is a polynomial correction,
$\epsilon_{top} \sim \exp(-\kappa l)\, l^\beta$. Eq.~(36) of \cite{RHG}
predicts such a correction and the numerical simulation finds it. However,
the exponents $\beta$ differ. Eq.~(36) of \cite{RHG} predicts $\beta
=-1/2$ for a strictly local error model. The numerical simulation
finds, for the close-to-local error model introduced above, $\beta
= -1.3 \pm 0.2$ in the time-like direction and $\beta = -0.9 \pm 0.2$ in
either space-like direction. 
.  Because of the uncertainty
in the values of $\beta$ we do not include the polynomial correction in
our analysis of the operational overhead. This is safe because it is to our
disadvantage. However, the exponential decay dominates and the effect
of the polynomial correction is very small.

The rate $\kappa$ of the dominant exponential decay of error is
potentially different along space-like and time-like directions, due
to the anisotropy of the error model. The numerical simulation finds
marginal differences at $p =p_c/3$,
\begin{equation}
\begin{array}{rclcl}
  \kappa &=& 0.85\pm 0.03&& \mbox{(time-like)},\\
  \kappa &=& 0.93\pm 0.03&& \mbox{(space-like)}.
\end{array}
\end{equation}

\paragraph{Error correction in $S$.} The $S$-qubits are involved in
  creating noisy ancilla states
$\rho^A\approx|A\rangle\langle A|$, $\rho^Y\approx |Y\rangle\langle
Y|$ encoded by the surface code, via the construction displayed in
Fig.~\ref{AP}. Due to the effective error on the $S$-qubits, these
ancilla states before distillation carry an error
$\epsilon^A_0:=1-\langle A|\rho^A|A\rangle$,
$\epsilon^Y_0:=1-\langle Y|\rho^Y|Y\rangle$ given by
\begin{equation}
  \label{IC}
  \epsilon^A_0= \epsilon^Y_0= 6p.
\end{equation}

\begin{figure}
  \begin{center}
    \includegraphics[width=10cm]{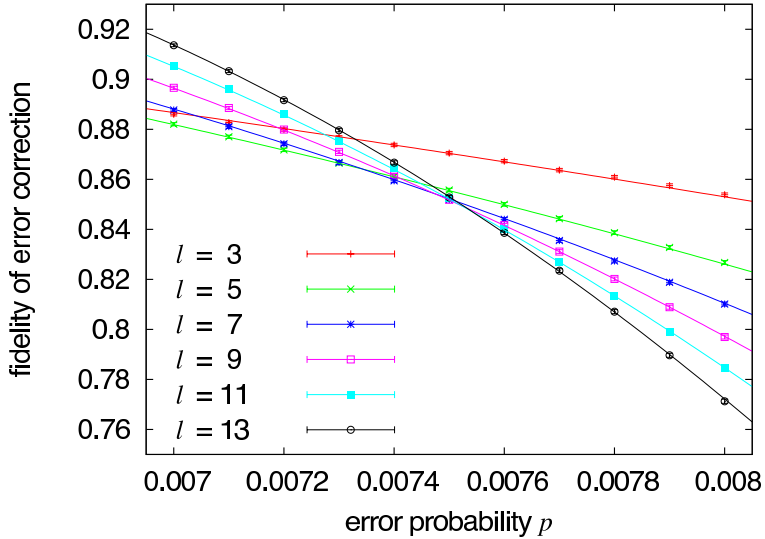}
    \caption{\label{Data}Numerical simulation for the topological
      threshold in $V$. The curves are best fits taking into account
      finite size effects of the lattice size $l$.  Beyond the smallest lattices,
      these finite size effects quickly vanish, and the curves
      intersect in a single point to a very good degree of
      accuracy. The value of $p$ at the intersection gives the threshold.}
  \end{center}
\end{figure}

\paragraph{Threshold.} There are two types of threshold within the
    cluster, the topological one in
$V$ and thresholds from $|A\rangle$ and $|Y\rangle$-state
distillation in $S$. An estimate $p_c^V$ to the topological
threshold is found in numerical simulation of finite-size lattices,
\begin{equation}
  \label{TopTh}
  p^V_c = 7.5 \times 10^{-3}.
\end{equation}
The result of the simulation is displayed in Fig.~\ref{Data}.

The recursion relations for state distillation, in the limit of
negligible topological error, are to lowest contributing order
$\epsilon^A_{l+1} = 35 (\epsilon^A_l)^3$ (c.f. \cite{BK04}) and
$\epsilon^Y_{l+1} = 7 (\epsilon^Y_l)^3$. The corresponding
distillation thresholds expressed in terms of the physical error rate
$p$ are
\begin{equation}
  p^A_c=\frac{1}{6\sqrt{35}} \approx 2.8\times 10^{-2},\;\;
  p^Y_c=\frac{1}{6\sqrt{7}} \approx 6.3\times 10^{-2}.
\end{equation}
The topological threshold is much smaller than the distillation
threshold, and therefore the former sets the overall threshold for
fault-tolerant quantum computation.

In our previous paper \cite{RHG}, the non-topological
threshold was the smaller one because the Reed-Muller quantum code was probed
in the error correction mode instead of the error detection mode
associated with state distillation \cite{BK04}. If we include
state-distillation into the setting of \cite{RHG}, the fault-tolerance
threshold increases to
\begin{equation}
\label{RHGth}
p_c=6.7\times 10^{-3},
\end{equation}
which is the topological threshold\footnote{The
value (\ref{RHGth}) differs from (\ref{TopTh}) due to minor
differences in the error model.
Specifically, \cite{RHG} requires a 2D structure of three or
more layers instead of a single one in the present
discussion. Consequently, the used operations and the order of
operations differ.}. The threshold (\ref{RHGth}) supersedes the
result of \cite{RHG}.

{\em{Remark 4.}} The effect of removing the redundant space-like oriented
$\Lambda(Z)$ gates (c.f. Remark 2) is to reduce the effective error on
the $S$-qubits. Eq. (\ref{IC}) then is replaced by $\epsilon_0^A =
\epsilon_0^Y = \frac{68}{15} p$. This affects neither the threshold
nor the overhead
scaling. The distillation threshold increases but it already
is the larger one. Also, as will be discussed in the next section, the
exponent which governs the overhead
scaling is a geometric quantity unaffected by the values of
$\epsilon_0^A$ and $\epsilon_0^Y$ in Eq. (\ref{IC}).

\section{Overhead}
\label{OH}

We are interested in the operational cost per gate, $O_3$, as a function of the
circuit size $\Omega$. To facilitate the calculation of $O_3$ it is helpful to
introduce the notions of the scale factor $\lambda$, the
defect thickness $d$, the gate length $L$, and
the gate volume $V$.

In the presented scheme, quantum gates are realized by twisting
defects. For that purpose alone, the defects could be line-like
structures. Then, the elementary cell of the lattice ${\cal{L}}$ constitutes a
building block out of which quantum gates and circuits are assembled.
However, in such a setting the property of
error correction is lost due to the presence of short error
cycles. To eliminate such errors the logical elementary cell
is rescaled to a cube of $\lambda \times \lambda \times
\lambda$ elementary cells. The cross-section of a defect with the
perpendicular plane becomes an area of $d \times d$ elementary cells (see Fig.~\ref{Cell2D}b).

The gate length $L$ is the total length of defect within a gate,
measured in units of the length of the logical cell. We will
subsequently use the gate length for an estimate of the gate error
remaining after topological error correction. The gate volume $V$ is
the number of logical cells that a gate occupies, each
consisting of $\lambda^3$ elementary cells of ${\cal{L}}$. Each
elementary cell is built with 16 operations.

Let $\epsilon_{top}(G, \lambda,d)$ be the probability of
failure for a gate $G$, as a function of the scale factor $\lambda$,
defect thickness $d$ and of its circuit layout. The
operational overhead $O_3(G)$ is then
\begin{equation}
  \label{OpOv}
  O_3(G) = 16 \lambda^3 V_G \exp\left(\epsilon_{top}(G,\lambda,d)
  \Omega\right).
\end{equation}
The exponential factor comes from the expected number of repetitions for a
circuit composed of $\Omega$ gates $G$. For a given $\Omega$,
the overhead should be optimized with respect to choosing $\lambda(\Omega)$ and
$d(\Omega)$.

\subsection{CSS-gates}

\begin{figure}[htb]
  \begin{center}
    \includegraphics[width=11cm]{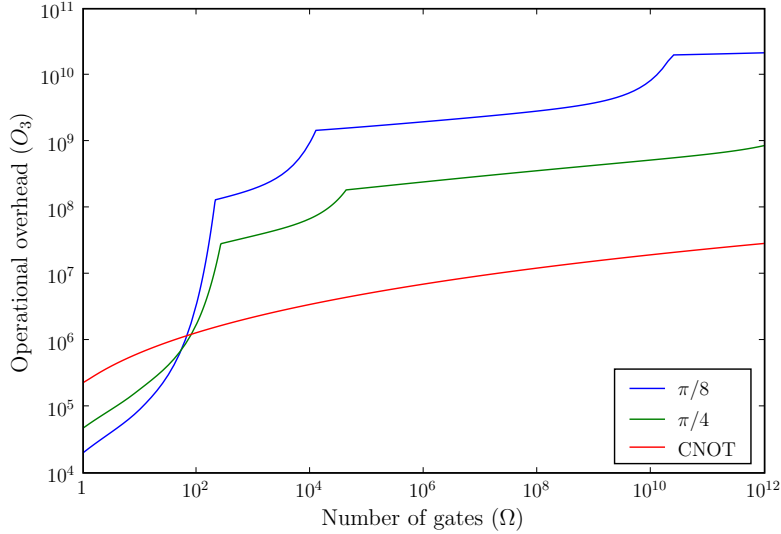}
    \caption{\label{O3}Operational overhead as a function of the
      circuit size. The upper curves are for the gates
      $\exp(i\frac{\pi}{8} Z)$ and  $\exp(i\frac{\pi}{4} X)$, the
      lower for the CNOT.}
  \end{center}
\end{figure}

The simplification for CSS gates is that no $S$-qubits are involved
and all operations are topologically protected.
To perform the optimization in Eq. (\ref{OpOv}) we need to know the
gate error $\epsilon_{top}$ as a function of $G$, $\lambda$, $d$. The errors
leading to gate failure may either be cycles wrapping around defects
of opposite color or relative cycles ending in defects of matching
color. The probability of gate failure is exponential in the length of the
shortest cycle or relative cycle, and proportional to the
number of such  error locations. The minimal cycle length is
$4(d+1)$ and the number of such cycles is equal to the gate length
$\lambda L_G$. The minimal length of a relative cycle leading to an error is
$\lambda-d$. It stretches between two neighboring defect segments one
logical cell apart. The number of such
relative cycles is at most $2 L_G \lambda (d+1)$. There are shorter relative
error cycles near junctions, but they are homologically equivalent to
the identity operation,
\begin{equation}
  \parbox{6cm}{\includegraphics[width=6cm]{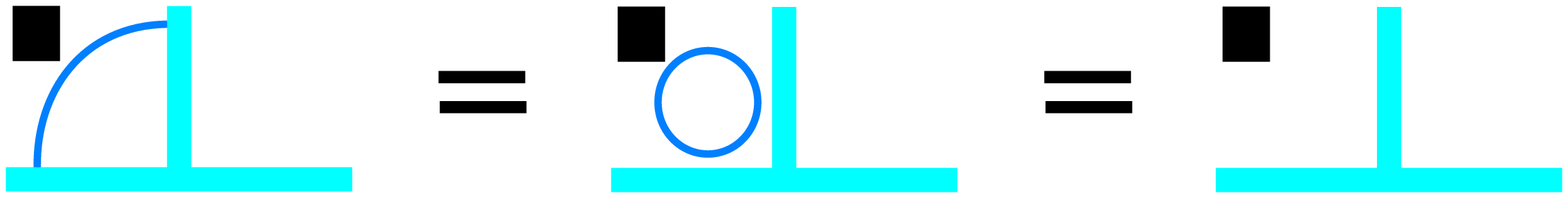}}
\end{equation}
Thus, the gate failure rate is
\begin{equation}
  \label{GF}
  \epsilon_{top}(L_G,\lambda,d) = \lambda
  L_G\left(\exp\left(-4\kappa\,(d+1)\right)+2
  (d+1)\exp\left(-\kappa(\lambda-d) \right) \right).
\end{equation}
We may now use this expression in Eq.~(\ref{OpOv}) and optimize for
given computational size $\Omega$. As an example, the operational
overhead for the CNOT-gate of Fig.~\ref{tg} is displayed in
Fig.~\ref{O3}.\medskip

\paragraph{The scaling limit.} We now perform the optimization of $O_3$
  in (\ref{OpOv}) with respect to
$\lambda$, in the limit of large circuit sizes $\Omega$. First, the
gate error $\epsilon_{top}$ in (\ref{GF}) is minimized when both exponentials
in (\ref{GF}) fall off equally fast, i.e. $d_{opt} = \lambda_{opt}/5$ for large
$d$, $\lambda$. Further, the overhead $O_3$  in (\ref{OpOv}) is minimized near
\begin{equation}
\label{OmEps}
\epsilon(\lambda(\Omega)) =  1/\Omega.
\end{equation}
Then $\lambda_{opt} \sim \ln\Omega/\kappa$, and
\begin{equation}
 \label{O23CSSscaling}
 O_3 \sim \frac{\ln^3\Omega}{\kappa^3}.
\end{equation}

\subsection{Non-CSS gates}

The estimation of the overhead for the non-CSS operation is along
the same lines but more complicated, due to the involved magic state
distillation. Every level $l$ of distillation is associated with its own
scale factor $\lambda_l$ and a defect thickness $d_l$. The
optimization of $O_3$ thus is over the larger set of parameters
$\Lambda = \{\{\lambda_0, d_0, \lambda_1, d_1, ... ,
\lambda_{l_{max}},d_{l_{max}}\}, l_{max}\}$.

Further, there are now two types of error. First the
previously discussed error of topological error correction resulting
from non-trivial error-cycles far away from the $S$-qubits. Second,
there is error associated with the $S$-qubits where topological
error correction breaks down.

The distillation of states $|A\rangle$ and $|Y\rangle$ uses
$S$-qubits at the lowest level. $|A\rangle$-distillation is based on
the $[15,1,3]$ Reed-Muller quantum code
\cite{BK04} and the distillation of $|Y\rangle$ on the $[7,1,3]$
Steane code. The $|A\rangle$-distillation
is performed using the circuit displayed in Appendix~\ref{RM}, of
volume $V_A$ and length $L_A$ (See Table~\ref{CircParam}). At each
level $l$ it requires 15 states $|A\rangle$ of level $l-1$ and, on average
$1705/512\approx 3.33$ states $|Y\rangle$. It succeeds with a probability of
$1-15\epsilon^A_{l-1}-\epsilon_{top}(L_A,\lambda_{l-1},d_{l-1})$, where
$\epsilon^A_{l-1}$ is the error of the states $|A\rangle$ at level
$l-1$. If successful, the residual error in the states $|A\rangle$ at
level $l$ is $\epsilon_l^A= 35 (\epsilon_{l-1}^{A})^3+\epsilon_{top}(L_Y,
\lambda_{l-1},d_{l-1})$. The distillation for $|Y\rangle$, of volume
$V_Y$ and length $L_Y$, takes 7 states $ |Y\rangle$ of the next-lower
level, and succeeds with a probability of
$1-7\epsilon^Y_{l-1}-\epsilon_{top}(L_Y,\lambda_{l-1},d_{l-1})$. If it
succeeds, the residual error is
$\epsilon_l^Y = 7
(\epsilon^Y_{l-1})^3+\epsilon_{top}(L_Y,\lambda_{l-1},d_{l-1}) $.

The above expressions for success probability and residual errors hold
to leading order in the contributing error probabilities
$\epsilon_{l-1}^{A}$, $\epsilon_{l-1}^{A}$. Further, a
gate error cannot simultaneously lead to termination of the circuit
and to a residual distillation error. Thus, we overestimate both
error probabilities by adding the full weight
$\epsilon_{top}(L,\lambda_{l-1},d_{l-1})$ to them.

The operational overheads for state distillation at level $l$,
$O^A_{3,l}$ and $O^Y_{3,l}$, and the corresponding residual errors
$\epsilon_l^A$, $\epsilon_l^Y$ are described by the recursion
relation
\begin{equation}
  \label{O3recrel}
  \begin{array}{rcl}
  O_{3,l}^A &=& \frac{1}{1-15\epsilon_{l-1}^A-\epsilon_{top}(L_A,
    \lambda_{l-1},d_{l-1})} \left(15\, O_{3,l-1}^A
    + \frac{1705}{512}\, O_{3,l-1}^Y + 16\lambda_{l-1}^3 V_A
    \right),\vspace{1mm}\\
  O_{3,l}^Y &=& \frac{1}{1-7\epsilon_{l-1}^Y-\epsilon_{top}(L_Y,
    \lambda_{l-1},d_{l-1})} \left(7\, O_{3,l-1}^Y
    + 16\lambda_{l-1}^3 V_Y \right),\vspace{1mm}\\
  \epsilon^A_l &=& 35 (\epsilon^A_{l-1})^3+
    \epsilon_{top}(L_A,\lambda_{l-1},d_{l-1}),\vspace{1mm}\\
  \epsilon^Y_l &=& 7 (\epsilon^Y_{l-1})^3+
    \epsilon_{top}(L_Y,\lambda_{l-1},d_{l-1}).
\end{array}
\end{equation}
The initial conditions are $O_{3,0}^A=O_{3,0}^Y=16$, and (\ref{IC}).

The distillation outputs states $|A\rangle$ and $|Y\rangle$ at level
$l_{max}$. One such state $|A\rangle$ and, on average, 1/2 state
$|Y\rangle$ is used to implement a gate
$\exp(i\frac{\pi}{8} Z)$ via the circuit displayed in
Fig.~\ref{gadgets}a, of volume $V_{1,z}$ and length
$L_{1,z}$. Its overhead is
\begin{equation}
  O_3^{\pi/8} = \left(O_{3,l_{max}}^A + \frac{1}{2} O_{3,l_{max}}^Y
  + 24(\lambda_{l_{max}})^3\, V_{1,z}\right) \exp\left(
  \left(\epsilon^A_{l_{max}}+ \epsilon^Y_{l_{max}} +
  \epsilon_{top}\left(L_{1,z}\lambda_{l_{max}}, d_{l_{max}}\right)
  \right)\Omega \right).
\end{equation}
The operational overhead needs to be optimized over the parameter set
$\Lambda$. This
has been done numerically \cite{Byrd}, and the result is shown in
Fig.~\ref{O3}.

\begin{table}
  \begin{center}
    \begin{tabular}{|l|r|r|} \hline \hline
      \multicolumn{1}{|c}{gate} & \multicolumn{1}{|c|}{volume} &
      \multicolumn{1}{|c|}{length}\\ \hline \hline
      CNOT-gate of Fig.~\ref{tg}a (packed)& $V_2=12$ & $L_2=22$ \\ \hline
      $U_Z$-gate of Fig.~\ref{gadgets}c & $V_{1,z}=2$ & $L_{1,z}=3$ \\ \hline
      $U_X$-gate of Fig.~\ref{gadgets}d & $V_{1,x}=4$ & $L_{1,x}=4$ \\ \hline
      $|Y\rangle$-distillation circuit & $V_Y=120$ & $L_Y=120$ \\ \hline
      $|A\rangle$-distillation circuit of Fig.~\ref{RMenc} & $V_A=336$
      & $L_A=362$\\
      \hline \hline
    \end{tabular}
    \caption{\label{CircParam}The gate volume and length for various
      gates and sub-circuits.}
  \end{center}
\end{table}

\paragraph{The scaling limit.} First, we compare the two contributions
to $\epsilon^A_l$ in (\ref{O3recrel}), $35(\epsilon^A_{l-1})^3$ and
$\epsilon_{top}$. If $\epsilon_{top}$ is much
larger than $35(\epsilon^A_{l-1})^3$ it inhibits the
convergence of ancilla distillation. Additional distillation rounds are needed
which are the most expensive component. If, to the contrary,
$\epsilon_{top}$ becomes much
smaller than $35(\epsilon^A_{l-1})^3$ it does not help the ancilla
distillation anymore but blows up the size of the logical
cell. Therefore, for optimal operational resources, both
contributions are comparable. Then, in the large size limit, $\ln
\epsilon^A_l = 3 \ln \epsilon^A_{l-1}$, $\lambda_l = 3 \lambda_{l-1}$,
$d_l =3 d_{l-1}$. Further, the success probabilities
$1-15\epsilon_{l}^A$ and $1-7\epsilon^Y_{l}$ for ancilla distillation
quickly approach unity with increasing distillation level
$l$. Therefore, in the large size limit, for the point of optimal operational
resources, the recursion relations (\ref{O3recrel}) can be replaced by
\begin{equation}
  \label{LinRec}
  \left(\begin{array}{c} O_3^A\\ O_3^Y \\ \lambda^3 \\ d \\ \ln
      \epsilon^A\\ \ln \epsilon^Y \end{array}\right)_l =
  \left(
    \begin{array}{cccccc}
       15 & \frac{1705}{512} & 16\,V_A & & & \\
       0 & 7 & 16\, V_Y & & & \\
       0 & 0 & 27\\
       & & & 3\\
       & & & & 3\\
       & & & & & 3
    \end{array}
  \right)
  \left(\begin{array}{c} O_3^A\\ O_3^Y \\ \lambda^3 \\ d \\ \ln
  \epsilon^A\\ \ln
  \epsilon^Y\\ \end{array}\right)_{l-1}
\end{equation}
Thus, $O^A_{3,l}, O^Y_{3,l} \sim 27^l$, $\ln \epsilon_l^A,
\ln \epsilon_l^Y \sim 3^l$. Then, with $\epsilon \sim 1/\Omega$ (\ref{OmEps}),
\begin{equation}
  \label{O3nonCSS}
  O^A_{3}, O^Y_{3} \sim (\ln \Omega)^3.
\end{equation}
Note that the distillation operations, for the case of perfect
CSS-gates, are associated with the more favorable scaling exponents
$\log_3 15\approx 2.46$ and $\log_3 7\approx 1.77$,
respectively. However, in our case the
topological error
protection of CSS gates must keep step with the rapidly decreasing
error of state distillation, by
adjusting the scale factor $\lambda$. This leads to a scaling exponent
of 3 for the CSS operational resources (c.f. Eq.~(\ref{O23CSSscaling})), which
dominates the resource
scaling of the entire state distillation procedure.

\paragraph{Discussion.} We have found that there is one dominant
exponent which governs the scaling of the operational overhead for
{\em{all}} gates from the universal set, $O_3 \sim \ln^3 \Omega$,
c.f. Eqs. (\ref{O23CSSscaling}) and
(\ref{O3nonCSS}). This exponent is a geometrical quantity. Its value, 3,
derives from the fact that the cluster state used in the scheme lives
in three spatial dimensions, and that errors are identified with
line-like objects (1-chains). Details of the implementation such as
the volume and length of the distillation circuits  play no role for
the scaling. This summarizes the main results of this section.

Let us now go beyond scaling and look at the pre-factors. Because of the
uniform overhead scaling the ratio of operational costs for non-CSS to
CSS-gates is constant in the limit of large computational size $\Omega$.
Inspection of Fig.~\ref{O3} shows that this ratio is in disfavor
of the non-CSS gates.

Without going into much detail, we would like to
point out that there is room for improvement here. The ratio
$O_3^A/O_3^{\Lambda(X)}$ is proportional to $V_A/{d^\prime}^3$. Herein,
$d^\prime$ is the shortest length of an error that goes undetected in
the distillation circuit. Its value is constrained by $1 \leq d^\prime
\leq d$, where $d$ is the distance of the used code (3 for the
Reed-Muller quantum code and the Steane code. The code distance $d$
shall not be confused with the defect thickness $d$ introduced earlier
in this section.). In the present
discussion, c.f. paragraphs preceding Eq.~(\ref{O3recrel}), we have used the
lower bound $d^\prime =1$ to simplify  the error counting.

The advise is to replace the Reed-Muller (Steane) quantum code by
another $[n,k,d]$ CSS-code for which the encoded gates $\exp(i\frac{\pi}{8}
\overline{Z}_i)$ ($\exp(i\frac{\pi}{4} \overline{Z}_i)$), $i=1\, ..\, k$, are
transversal, and
with the additional properties of having a large distance $d$ and a
good ratio $k/n$. Such codes need to be searched for systematically.

The distillation circuits can be further optimized. The logical circuit depth
can be reduced to 3 for $|A\rangle$-distillation and 2 for
$|Y\rangle$-distillation, independent of
the the code parameters $n$, $k$ and $d$. The circuit height 3 remains
unchanged. Thus, the circuit volumes per output qubit $V_A$, $V_Y$ can
be reduced to $V_A = 9 (n/k+1)$, $V_Y = 6 (n/k+1)$.

\section{Summary and outlook}
\label{OL}

In this paper we have discussed in detail the error threshold and
overhead for universal fault-tolerant quantum computation based
on the one-way quantum computer with a three-dimensional cluster state.
By conversion of one spatial cluster dimension into time we have
reduced the dimensionality of the scheme to two. Also, the described
scheme only requires translation-invariant nearest-neighbor Ising
interaction among the qubits. These features should facilitate future
implementation. We envision cold atoms in optical lattices \cite{OL1,OL2},
two-dimensional ion traps \cite{ITr}, quantum dot systems \cite{Qdot} and
arrays of superconducting qubits \cite{SCQ} as suitable candidate
systems for experimental realization.

On a more abstract level, we have initiated the discussion of the topological
properties of the defect configurations. We have described a set of
transformation rules that allow us to switch between equivalent
configurations. We have applied these rules to simplify sub-circuits
and to derive circuit identities, by a sequence of operations
reminiscent of the Reidemeister moves for link diagrams.

There are a host of questions that remain open, from the applied to
the abstract. Below a few are listed.
\begin{itemize}
\item{Optimization of the error threshold. With the current implementation of
    error correction we have exhausted the capabilities of the
    minimum-weight chain matching algorithm. There is one improvement
    that promises a noticeable gain. So far, error corrections
    on the mutually dual lattices ${\cal{L}}$ and $\overline{\cal{L}}$
    run entirely separate. However, errors on ${\cal{L}}$ and
    $\overline{\cal{L}}$ are correlated such that error correction
    could benefit from cross-talk between the two lattices.}
\item{Transversality of encoded gates. Motivated in part by the
    discussion of overhead in
    Section~\ref{OH} but a topic of more general theoretical
    interest is to find further stabilizer codes that posses the
    capability of performing non-Clifford gates transversally.}
\item{Robustness of error threshold. Here we have discussed a logic
    gate-based error model. What about more physical error models such
    as, for example, spins coupled to an Ohmic bath?}
\item{Connection with the category-theoretic work of Abramsky and
    Coecke. In this paper we have used an encoding with two holes
    per logical qubit. There is another encoding that gets by with a single
    hole, making additional use of the external system
    boundary. In that other code, for both the primal and the dual
    defects, the ``cups'' of Fig.~\ref{tg}b denote the preparation of
    a Bell state, and the corresponding ``caps'' Bell
    measurement. They provide a concrete physical realization of the
    corresponding abstract elements in the category-theoretic calculus
    of \cite{Coecke1}. Also, the authors of \cite{Coecke1} introduce a
    ``line of information flow''. It is represented by the defect
    strands in our scheme.

  If the teleportation identity was the only phenomenology supported
  by the defects we would not get very far in terms of fault-tolerant
  quantum computation. To this end, it is crucial to have two distinct
  types of qubits, primal and dual, which interact in a non-trivial
  manner (\ref{mono}). Now, the question is whether this enlarged
  phenomenology can be included in the category-theoretic framework of
  \cite{Coecke1} and whether it enriches that framework.}
\end{itemize}

\paragraph{Acknowledgments.} RR would like to thank Hans-Peter
  B{\"u}chler, Jiannis Pachos, Almut Beige, Simon Benjamin, Parsa Bonderson, Trey
  Porto and David Weiss for
  discussions. KG is supported by DOE Grant No. DE-FG03-92-ER40701. JH
  is supported by DTO. RR is supported by the Government of Canada
  through NSERC and by the Province of Ontario through MEDT.

\appendix

\section{Circuit for state distillation}
\label{RM}

\begin{figure}[h]
  \begin{center}
    \includegraphics[height=9cm]{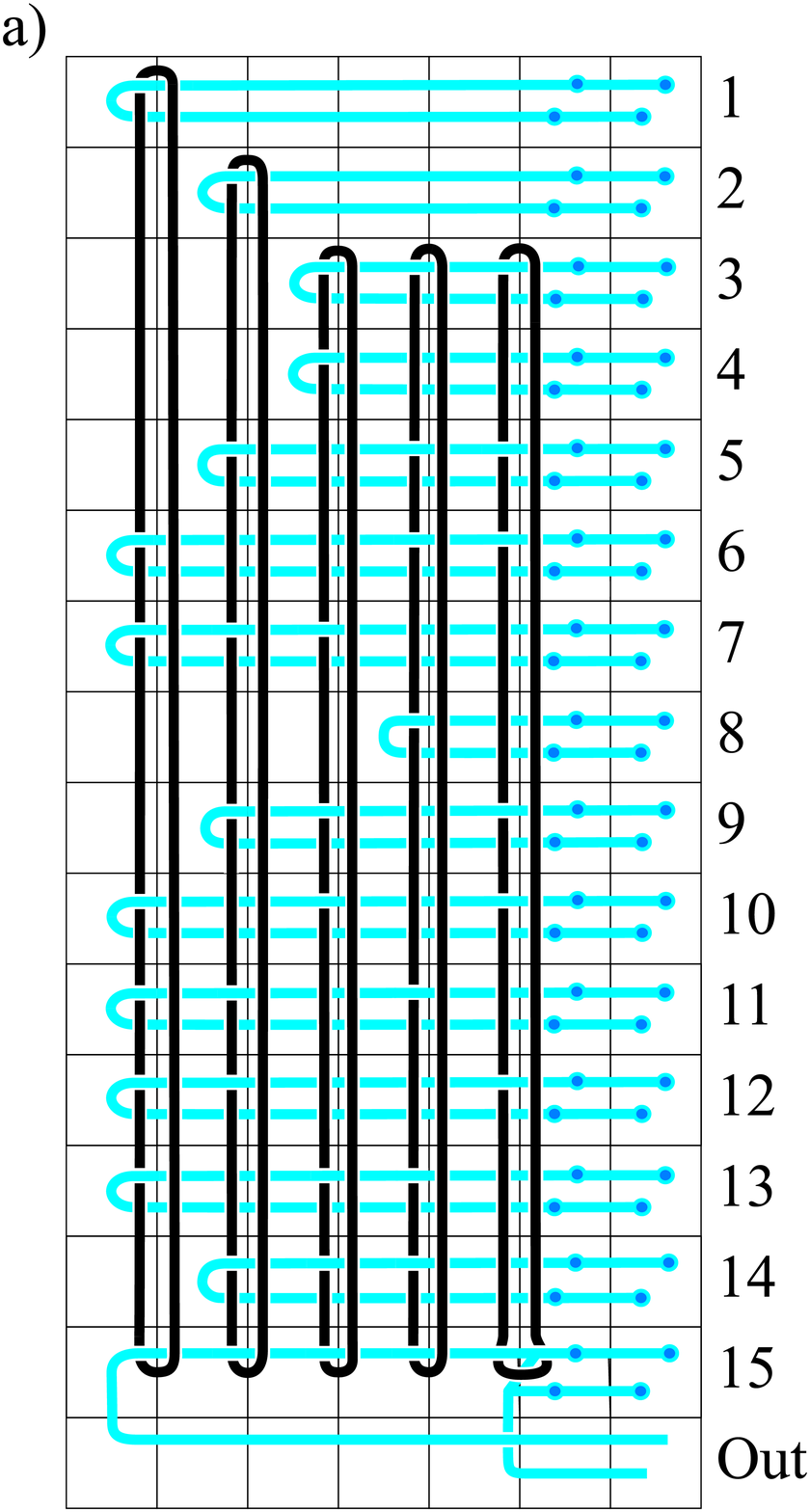}\hspace*{3mm}
    \includegraphics[height=9cm]{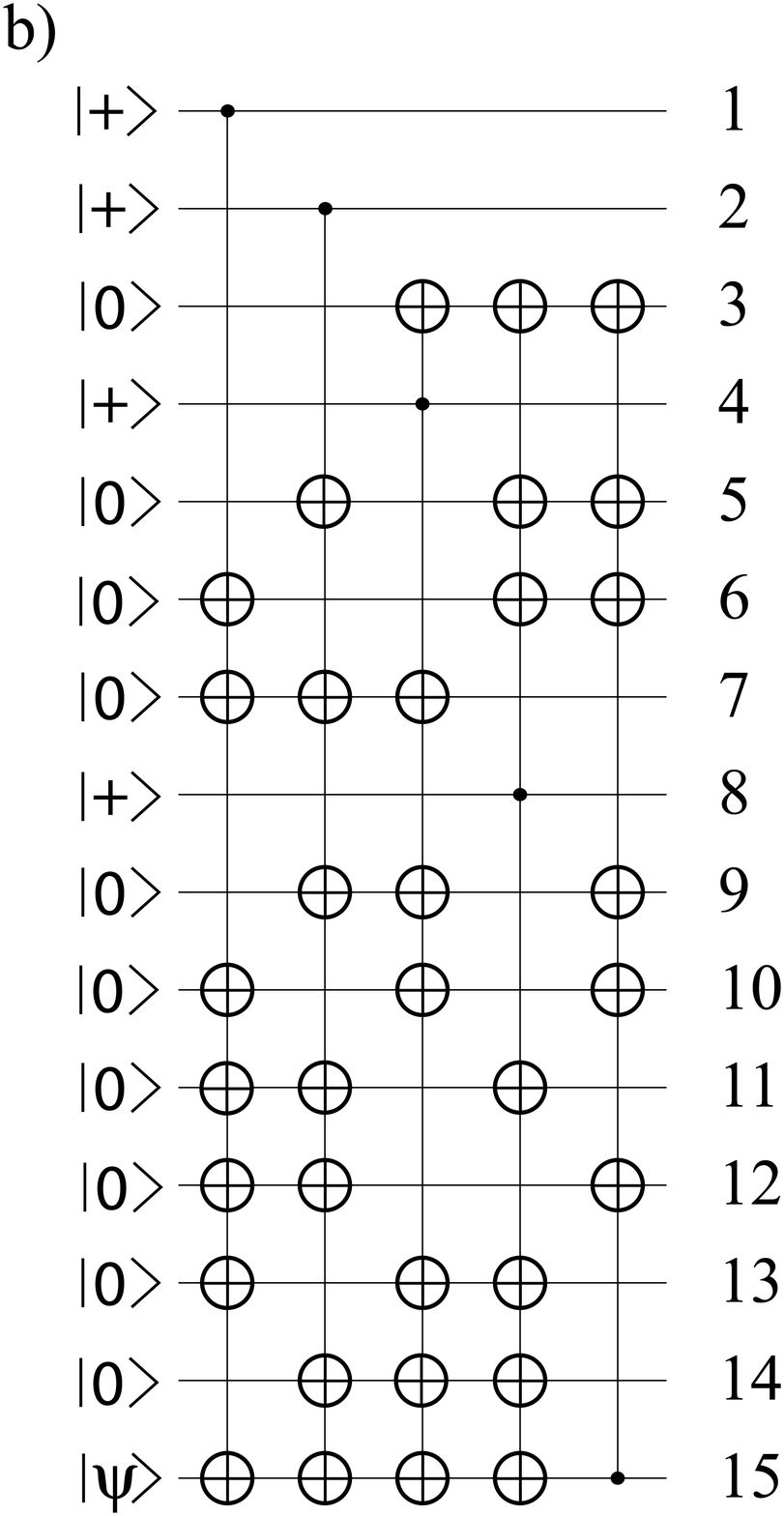}
    \caption{\label{RMenc}\QCcns-realization of the
    $|A\rangle$-state distillation (a). The dots on the defect lines
    are the ports to connect $|A\rangle$- and $|Y\rangle$ states for
    $Z$-rotations (c.f. Fig.~\ref{gadgets}). The main part in the
    distillation is the encoder
    for the $[15,1,3]$ Reed-Muller quantum code, displayed as a quantum
    circuit in (b).}
  \end{center}
\end{figure}

We use a variant of the magic state distillation circuit described in
\cite{BK04}, adapted to the \QCcns. The topological circuit is
displayed in Fig.~\ref{RMenc}. The
procedure is this: we start with a Bell state, encode one of its
qubits with the Reed-Muller quantum code and measure each of the 15
qubits leaving the encoder in the eigenbasis of $X_i - Y_i$.

The Reed-Muller quantum code \cite{BK04} has the nice property that the
$X$-syndrome can be measured in the $X$, $Y$, $X+Y$ and
$X-Y$-basis. Further,
$\frac{\overline{X} + \overline{Y}}{\sqrt{2}} = \bigotimes_{i=1}^{15}
\frac{X_i - Y_i}{\sqrt{2}}$. Therefore, through the local
$X-Y$-measurements and classical post-processing we can both learn
the $X$-syndrome and project the encoded qubit of the Bell pair into
an eigenstate of $\overline{X} + \overline{Y}$. Thus, we
simultaneously project the unencoded qubit into the state
$X^aZ^b|A\rangle$, with $a,b \in \{0,1\}$ depending on the measurement
outcomes and on which of the four Bell states was used. We keep this
qubit if the above $X$-syndrome measurements yield a trivial
outcome. In this case, the residual error $\epsilon_l$ is, to
leading order, $\epsilon_l^A=35(\epsilon_{l-1}^A)^3$ (c.f. \cite{BK04}).

The local $X-Y$-measurements are performed by a unitary operation
$\exp(-i\frac{\pi}{8} Z_i)$ followed by an $X_i$-measurement. Each such
unitary requires one ancilla $|A\rangle$ and, with probability 1/2,
one additional ancilla $|Y\rangle$ such that one round of magic state
distillation performed in this way consumes 15 states $|A\rangle$
and, on average, 15/2 states $|Y\rangle$. With a small
modification\footnote{Denote by ${\cal{J}}$ the set of
subsets of $\{1,2, .. , 15\}$ such that $\bigotimes_{i\in J}X_i$ is an
encoded gate (including the identity operation) for all $J \in
{\cal{J}}$. Then, for the
Reed-Muller quantum code, the Clifford unitary $\bigotimes_{i\in
  J}\exp(i\frac{\pi}{4} Z_i)$ is also an encoded gate $\forall\,J\in
{\cal{J}}$, namely $I$ or $\exp(-i\frac{\pi}{4} \overline{Z})$. Now
suppose that
after probabilistic implementation of the $\pi/8$-phase gates,
$\pi/4$-phase gates on a set
$K\subset\{1,2,..,15\}$ are required. Then, it is equivalent to
perform $U_{K\oplus J} = \bigotimes_{i \in
  K}\exp(i\frac{\pi}{4} Z_i) \bigotimes_{i \in J}\exp(i\frac{\pi}{4}
  Z_i)$, $\forall J \in {\cal{J}}$, modulo local
Pauli operators $Z_i$. (Note that $\exp(i\frac{\pi}{4} Z)|A\rangle = X
|A\rangle \cong |A\rangle$.) We minimize the support of $U_{K\oplus
  J}\, \mbox{mod}\, \{Z_i\}$ by varying $J\in
{\cal{J}}$. In this way, we reduce the average number of
$|Y\rangle$-states required in a distillation step to $1705/512\approx
3.33$.},
we can reduce the average number of required $|Y\rangle$-states to 1705/512.

The distillation circuit for $|Y\rangle$-states is constructed in a
similar manner. It is based on the Steane code and
requires seven input states $|Y\rangle$ in each round.

\section{Effective error model on ${\cal{L}}$ and
    $\overline{\cal{L}}$}

\begin{figure}
\begin{center}
  \includegraphics[width=5.5cm]{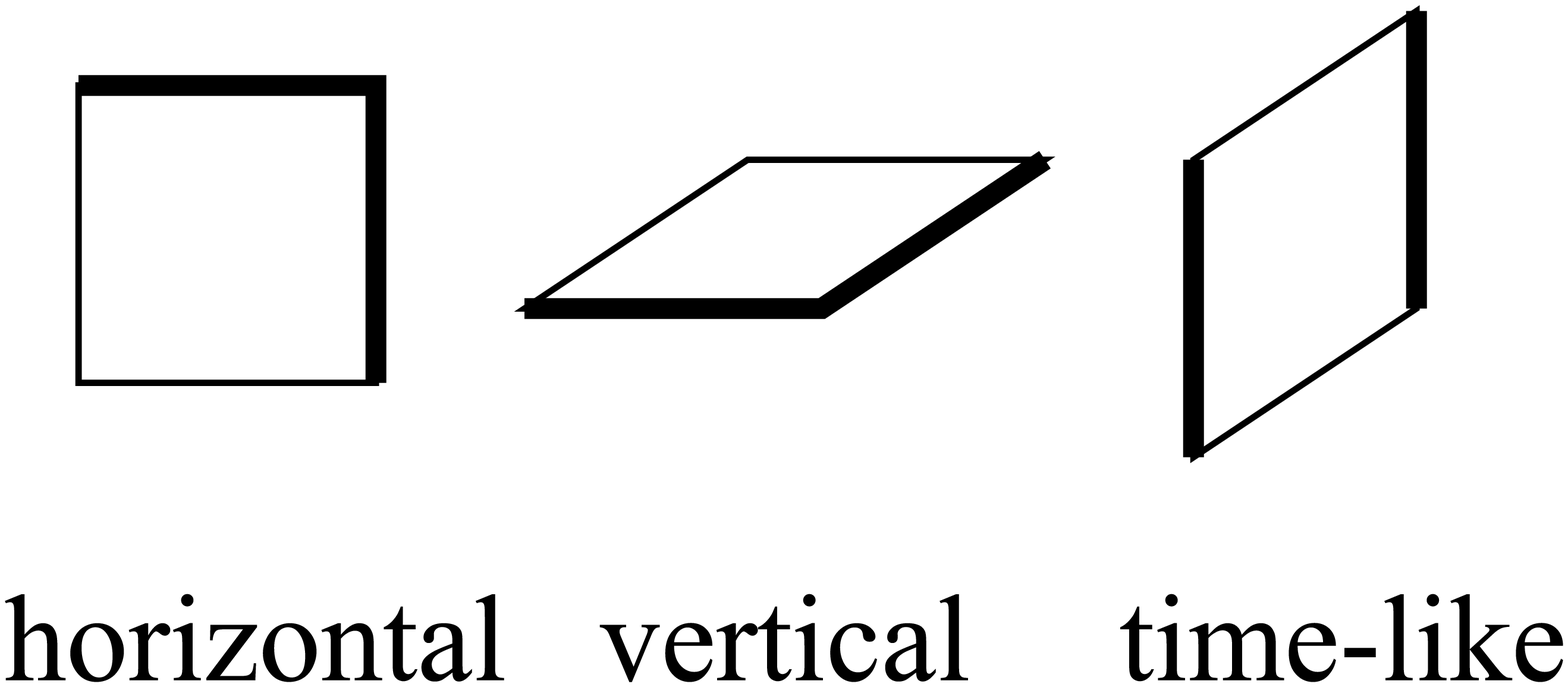}
  \caption{\label{CE} The correlated errors (same for on the primal and dual
    lattice). Shown are horizontal, vertical and time-like faces of
    ${\cal{L}}$ and $\overline{\cal{L}}$. Thick lines
    indicate error locations.}
\end{center}
\end{figure}

After the mapping described in Section~\ref{2D} the physical setting
is in two dimensions. However, the topological error correction is
still performed on the three-dimensional lattices ${\cal{L}}$ and
$\overline{\cal{L}}$. The error model of Section~\ref{FTT}, including
gate error for one and two-qubit gates, preparation and measurement,
effectively results in $Z$-errors on individual edges and correlated
errors on two edges of ${\cal{L}}$ or $\overline{\cal{L}}$,
respectively. The location of correlated errors is shown in
Fig.~\ref{CE}.

The effective error channel on time-like edges is
\begin{equation}
  \begin{array}{rcl}
  {\cal{T}}_{1,t}&=&\left(\left(1 - \frac{8}{15}p_2\right)[I_t] +
    \frac{8}{15} p_2 [Z_t] \right)^{\circ 2} \circ
  \left(\left(1 - \frac{2}{3}p_P\right)[I_t] +
    \frac{2}{3} p_P [Z_t] \right) \circ\\
  &&\left(\left(1 - \frac{2}{3}p_M\right)[I_t] +
    \frac{2}{3} p_M [Z_t] \right).
  \end{array}
\end{equation}
The effective error channel on a space-like edge---horizontal or
vertical---is
\begin{equation}
  \begin{array}{rcl}
  {\cal{T}}_{1,s}&=&\left(\left(1 - \frac{8}{15}p_2\right)[I_s] +
    \frac{8}{15} p_2 [Z_s] \right)^{\circ 3} \circ
  \left(\left(1 - \frac{2}{3}p_1\right)[I_s] +
    \frac{2}{3} p_1 [Z_s] \right)^{\circ 2}.
  \end{array}
\end{equation}
The effective error channel for each of the correlated errors
displayed in Fig.~\ref{CE} is
\begin{equation}
  \begin{array}{rcl}
  {\cal{T}}_2&=&\left(\left(1 - \frac{8}{15}p_2\right)[I_{ab}] +
    \frac{8}{15} p_2 [Z_a Z_b] \right).
  \end{array}
\end{equation}
The probability of time-like and space-like individual errors is
different. In the error correction procedure this is accounted for
by using non-uniform weights in the minimum-weight chain matching
algorithm \cite{AlgoEff}. Likewise, the correlated errors in the boundary of
space-like faces are accounted for by including additional diagonal
edges.

\end{document}